\documentclass[prd,aps,nofootinbib,floatfix,10pt]{revtex4}
\usepackage{amsmath,graphicx,epsfig,amssymb}
\usepackage{inputenc}
\usepackage[usenames]{color}
\usepackage{ulem} 
\usepackage{bigstrut}
\usepackage{slashed}
\usepackage{multirow}
\usepackage{subfigure}

\allowdisplaybreaks

\newcommand{\red}{\textcolor{red}}

\begin{document}
\title{Global Analysis of hadronic two-body $B$ decays in the perturbative QCD approach}
	
\author{Jun Hua$^{1}$~\footnote{Email: huajun$\underline{\;}$phy@sjtu.edu.cn},
Hsiang-nan Li$^{2}$~\footnote{Email: hnli@phys.sinica.edu.tw},
Cai-Dian L$\ddot{u}^{3,4}$~\footnote{Email: lucd@ihep.ac.cn },
Wei Wang$^{1}$~\footnote{Email: wei.wang@sjtu.edu.cn}, and
Zhi-Peng Xing$^{1}$~\footnote{Email: zpxing@sjtu.edu.cn}}
\affiliation{$^{1}$ INPAC, Key Laboratory for Particle Astrophysics and Cosmology (MOE), Shanghai Key Laboratory for Particle Physics and Cosmology,
		School of Physics and Astronomy, Shanghai Jiao-Tong University, Shanghai
		200240, China }
\affiliation{$^{2}$Institute of Physics, Academia Sinica,
                 Taipei, Taiwan 11529, Republic of China }
\affiliation{$^{3}$ Institute of High Energy Physics, Chinese Academy of Sciences,
                 Beijing 100049, China }
\affiliation{$^{4}$   School of Physics, University of Chinese Academy of Sciences, Beijing 10049, China}

\begin{abstract}
Based on the flavor structure of four-quark effective operators,
we develop an automatic computation program to calculate hadronic two-body $B$ meson decay amplitudes,
and apply it to their global analysis in the perturbative QCD (PQCD) approach. Fitting the PQCD
factorization formulas for $B\to PP,VP$ decays at leading order in the strong coupling $\alpha_s$
to measured branching ratios and direct CP asymmetries, we determine the Gegenbauer moments in
light meson light-cone distribution amplitudes (LCDAs). It is found that most of the fitted
Gegenbauer moments of the twist-2 and twist-3 LCDAs for the pseudoscalar meson $P$ ($P=\pi$, $K$)
and vector meson $V$ ($V=\rho$, $K^*$) agree with those derived in QCD sum rules. The shape
parameter for the $B_s$ meson distribution amplitude and the weak phase $\phi_3(\gamma)=(75.2\pm2.9)^\circ$
consistent with the value in Particle Data Group are also obtained.
It is straightforward to extend our analysis to higher orders and higher powers in
the PQCD approach, and to the global determination of LCDAs for other hadrons.
\end{abstract}
\maketitle
	
\section{Introduction}

The study of  heavy quark physics is firmly  in the precision era nowadays. On the
experimental side, the $B$ factories, {\it i.e.}, the BaBar, Belle,  and LHCb  have
collected abundant data of exclusive $B$ meson decays, which can be employed not only
to explore involved rich QCD dynamics, but also to probe the origin of CP violation and potential
new physics signals~\cite{Aoki:2019cca}. Vastly  more data will be still accumulated by the upgraded LHCb and
Belle-II Collbarotions~\cite{Bediaga:2012py,Kou:2018nap,Cerri:2018ypt}.
On the theoretical side, tremendous progress on the development of QCD treatments of exclusive $B$ meson decays with
controllable uncertainties  has been achieved.  Strict confrontation between data and theoretical expectations
has led to some mild tensions between experimental observations and the Standard Model~\cite{Aoki:2019cca},
which may be vaguely attributed to  new physics beyond the Standard Model.
This undoubtedly motivates the attempt  to gain deeper understanding of QCD dynamics in exclusive
$B$ meson decays and better control of hadronic uncertainties.

The $b$ quark mass $m_b$ is much larger than the QCD hadronic scale $\Lambda_{\rm QCD}$, which
renders QCD analyses of exclusive $B$ meson decays possible. Nonperturbative dynamics
in heavy meson decays is reflected by infrared divergences in radiative corrections.
When a factorization theorem holds,
infrared divergences are absorbed into hadron light-cone distribution amplitudes (LCDAs),
so that the remnant, being infrared finite, is calculable at the parton level in perturbation theory.
A physical quantity, such as a heavy-to-light transition form factor, is then factorized into a
convolution of a $b$ quark  decay hard kernel with hadron LCDAs in parton momentum fractions.
The corresponding factorization theorem should be proved to all orders in the strong
coupling $\alpha_s$ and to certain power in $\Lambda_{\rm QCD}/m_b$. LCDAs,
despite of being nonperturbative, are universal, {\it i.e.}, process-independent.
With this universality, LCDAs, determined by nonperturbative methods like QCD sum rules~\cite{Ball:2004ye, Ball:2006wn,Ball:2007rt}
and lattice QCD~\cite{Bali:2017ude,Bali:2019dqc,Hua:2020gnw}, or extracted from experimental data, can be employed
to make predictions for other modes involving the same hadrons.

The theoretical approaches based on factorization theorems in the heavy quark limit
include light-cone QCD sum rules (LCSR) \cite{Chernyak:1990ag,Ali:1993vd,Khodjamirian:2000mi},
the QCD-improved factorization (QCDF) \cite{Beneke:1999br},
the perturbative QCD (PQCD) factorization \cite{Li:1994cka,Li:1995jr,Li:1994iu,Keum:2000ms,Keum:2000wi,Lu:2000em},
and the soft-collinear effective theory (SCET) \cite{Bauer:2000ew,Bauer:2000yr}. The collinear
factorization applies to relevant correlators in LCSR, where some hadronic states are
expanded into parton Fock states characterized by different twists.
The QCDF approach is an extension of the naive factorization assumption for
hadronic two-body $B$ meson decays in the collinear factorization theorem. The SCET for kinematic
regions with energetic final state hadrons is equivalent to the collinear
factorization theorem, but formulated in terms of effective operators. The $k_T$
factorization theorem is the basis of the PQCD approach, which is more appropriate in
the endpoint region of parton momentum fractions.
Many efforts have been devoted to systematic investigation of hadronic two-body $B$ meson decays at
various orders in $\alpha_s$ and powers in $\Lambda_{\rm QCD}/m_b$ ~\cite{Li:2020rcg,Huber:2016xod,Bell:2020qus}.
In all the above formalisms nonperturbative hadron LCDAs provide a major source of theoretical uncertainties.

A hadron LCDA can be expanded into a series of Gegenbauer polynomials with the coefficients,
namely, the Gegenbauer moments being determined by other methods
as aforementioned. Though some attempts have been made to calculate Gegenbauer moments using lattice
QCD~\cite{Bali:2017ude,Bali:2019dqc,Hua:2020gnw}, not all LCDAs are constrained in this way so far.
Here we will perform a global fit of the Gegenbauer moments in light meson LCDAs to measured branching ratios
and direct CP asymmetries in hadronic two-body $B$ meson decays  in the PQCD approach.
The decay amplitudes at leading order (LO) of the strong coupling $\alpha_s$
will be  constructed automatically with a computation program  by making use of flavor SU(3) properties.
We establish a Gegenbauer-moment-independent database, by means of which each decay amplitude
is expressed as a combination of the relevant Gegenbauer moments and
Cabibbo-Kobayashi-Maskawa (CKM) matrix elements. The Gegenbauer moments in the
leading-twist (twist-2) and next-to-leading-twist (twist-3) LCDAs for the pseudoscalar meson $P$
($P=\pi$, $K$) and vector  meson $V$ ($V=\rho$, $K^*$) are then fixed in the global fit,
most of which are found to agree with those from QCD sum rules
\cite{Ball:2004ye,Ball:2006wn,Ball:2007rt}. It should be noticed that the precision of extracted LCDAs
depends on the accuracy of the involved hard kernels.  As a by-product, the shape
parameter for the $B_s$ meson distribution amplitude (DA) and the weak phase $\phi_3(\gamma)=(75.2\pm2.9)^\circ$
consistent with the value in Particle Data Group \cite{Zyla:2020zbs} are also obtained.
Though we have focused on the $B\to PP,VP$ decays at LO in PQCD,
our work provides the first setup for a global analysis of exclusive $B$ meson decays, and can be
generalized straightforwardly to include other modes, and higher-order and/or higher-power corrections.

The rest of this paper is organized as follows. We give a brief overview of the theoretical
framework for hadronic two-body $B$ meson decays in Sec.~II. The automatic derivation of the decay
amplitudes in the PQCD approach is formulated in Sec.~III, where the Gegenbauer-moment-independent database
for the considered modes is established. We perform a global fit of meson LCDAs and the CKM angle $\phi_3(\gamma)$ to
a limited number of physical observables in the $B\to PP,VP$ decays, and present the numerical results  in Sec.~IV.
We also compare our predictions for some other modes excluded in the fit
with experimental data. A few remarks and future improvements on our analysis are outlined
at the end of this section. Section V contains a summary of the present work. The explicit
factorization formulas and their ingredients are  collected  in the Appendix.

\section{THEORETICAL FORMALISM}

In exclusive processes, such as heavy-to-light transition form factors,
the range of a parton momentum fraction $x$, contrary to that in an inclusive case, is
not experimentally controllable, and runs from 0
to 1. Hence, the endpoint region with  $x\to 0$ is unavoidable. If
no endpoint singularity is developed, implying that the endpoint region is likely
 power suppressed, the collinear factorization will work. If
such a singularity occurs, the collinear
factorization will break down, and the $k_T$ factorization should be adopted. In fact,
the observation $QF_2(Q^2)/F_1(Q^2)\sim$ const. \cite{Jones:1999rz,Gayou:2001qd},
$F_1$ and $F_2$ being the proton Dirac and Pauli form factors,
respectively,  and $Q$ being a momentum transferred, indicates that the $k_T$ factorization
is an appropriate tool for studying exclusive processes \cite{Ralston:2003mt}. It has been shown
that infrared divergences appearing in loop corrections to exclusive processes can be absorbed
into hadron LCDAs in the $k_T$ factorization without breaking the gauge invariance \cite{Nandi:2007qx}.
Since the  $k_T$ factorization theorem
was proposed \cite{Botts:1989kf,Li:1992nu}, there had been broad applications to various
processes \cite{Li:2001ye}.

The application of the collinear factorization theorem to exclusive $B$
meson decays, for instance, the $B\to\pi$ transition form factors, suffers the endpoint singularities
mentioned above \cite{Szczepaniak:1990dt,Burdman:1992hg,Beneke:2000wa}: the twist-2 and
twist-3 contributions are logarithmically and linearly divergent, respectively.
The inclusion of parton transverse momenta $k_T$, regulating the endpoint
singularities, induces soft logarithms in higher-order corrections. Their overlap with the existent
collinear logarithms generates the double logarithms $\alpha_s\ln^2 k_T$,
which must be organized in order not to spoil perturbative expansion.
The basic idea for the $k_T$ resummation of the double
logarithms into a Sudakov factor has been elaborated in
\cite{Botts:1989kf,Collins:1981uk,Li:1994cka,Li:1995jr,Li:1994iu}, where
the explicit expressions of the Sudakov exponents can be found. The resultant Sudakov
suppression on the low $k_T$ contribution in the endpoint region renders the magnitude of $k_T^2$
roughly $O(m_b\Lambda_{\rm QCD})$. The coupling constant
$\alpha_s(\sqrt{m_b\Lambda_{\rm QCD}})/\pi \sim 0.13$ is
then small enough to justify the perturbative evaluation of heavy-to-light transition form
factors at large recoil \cite{Keum:2000ms,Kurimoto:2001zj,Wei:2002iu}.

On the other hand, the double logarithms $\alpha_s\ln^2 x$ from radiative
corrections were observed in the semileptonic decay
$B\to\pi l\nu$ \cite{Akhoury:1993uw} and in the radiative decay $B\to\gamma l\nu$
\cite{Korchemsky:1999qb}. It has been argued that when the endpoint region is
important, these double logarithms should be organized into a quark jet function systematically
in order to improve perturbative expansion. The procedure is referred
to as the threshold resummation \cite{Li:2001ay}. The resultant jet function has been
shown to vanish quickly as $x\to 0$. It turns out that in a
self-consistent perturbative evaluation of the heavy-to-light transition form factors,
where the original factorization formulas are further convoluted with the jet
function, the endpoint singularities do not exist \cite{Li:2001ay}.
The threshold resummation for the jet function
has been pushed to the next-to-leading-logarithm accuracy recently \cite{Zhang:2020qaz}.
Note that either the threshold or $k_T$ resummation smears the endpoint
singularities. To suppress the soft contribution sufficiently,
both resummations are required, such that reliable results for the
heavy-to-light transition form factors can be attained.

We emphasize that the power counting for a parton transverse momentum $k_T$ is nontrivial,
compared to the power counting for the fixed scales like $m_b$ and $\Lambda_{\rm QCD}$.
The $k_T$ factorization is suitable for a multi-scale process, like a heavy-to-light transition
form factor, to which the region of a small momentum fraction $x$ dominates.
The small $x$ introduces an additional intermediate scale
$xm_b^2\sim m_b\Lambda_{\rm QCD}$, respecting the hierarchy $m_b^2\gg xm_b^2\gg\Lambda_{\rm QCD}^2$.
A parton $k_T$, being an integration variable in a $k_T$ factorization formula, can take values of orders
of the above scales. The $k_T$ factorization should apply, as a hard kernel
depends on the large scale $m_b^2$ and the intermediate scale $m_b\Lambda_{\rm QCD}$,
but not on the small scale $\Lambda_{\rm QCD}^2$, and the factorization of hadron wave
functions hold for a parton $k_T$ at both the intermediate and small scales. Once these criteria
are satisfied, the $k_T$ dependence in a hard kernel is not negligible \cite{Nandi:2007qx},
and a convolution between the hard kernel and the wave functions in $k_T$ is demanded.
If a hard kernel involves only the large scale, the $k_T$ dependence of the hard kernel can be
neglected. It is then integrated out in the wave functions, and one is led to the collinear factorization.


Since a wave function contains the contributions characterized by both the intermediate and small
scales, it is legitimate to further factorize the former out of the wave function, as the intermediate scale
is regarded as being perturbative. This gives the aforementioned $k_T$ resummation, which is justified
perturbatively for the scale $k_T^2\sim m_b\Lambda_{\rm QCD}$.
After this organization, the remaining piece, ie., the initial condition for the Sudakov
resummation, involves only the small scale $\Lambda_{\rm QCD}^2$, and corresponds to
a hadron DA. Note that a more sophisticated
formalism, called the joint resummation, which organizes
the mixed logarithms formed by the above two different scales, has been developed in \cite{Li:2013xna}.
Similarly, it is also legitimate to further factorize the contribution characterized by
an intermediate scale out of a hard kernel in the $k_T$ factorization. This re-factorization
yields the jet function, through which the logarithms of $xm_b^2$
are resummed to all orders.

The effective Hamiltonian for hadronic two-body $B$ meson decays is given by
\begin{eqnarray} \label{eq:hamiltonian}
  {\cal H}_{eff} &=& \frac{G_{F}}{\sqrt{2}}
     \bigg\{ \sum\limits_{q=u,c} V_{qb} V_{qD}^{*} \big[
     C_{1}  O^{q}_{1}
  +  C_{2}  O^{q}_{2}\Big]- V_{tb} V_{tD}^{*}{\sum\limits_{i=3}^{10}} C_{i}  O_{i}\bigg\}+ \mbox{H.c.} ,
\end{eqnarray}
with the Fermi constant $G_F$, the CKM matrix elements $V_{qb(D)}$, $D=d,s$,
the local four-quark operators $O_i$, and the Wilson coefficients $C_i$.
All the factorizable, nonfactorizable and power-suppressed annihilation contributions resulting from
the above four-quark operators are calculable in the PQCD approach without the endpoint singularities.
The arbitrary cutoffs introduced in QCDF \cite{Beneke:2000ry,Beneke:2001ev} are not necessary, and
PQCD factorization formulas involve only universal and controllable inputs.
The $B\to M_2M_3$ decay amplitude is generically  factorized
into the convolution of the Wilson coefficient $C$, a six-quark hard kernel $H$,
the jet function $J_t$, and the Sudakov factor $S$
with meson LCDAs $\phi$ \cite{Chang:1996dw,Yeh:1997rq,Cheng:1999gs,Li:2001vm},
\begin{eqnarray}\label{eq:six}
A=\phi_B\otimes C\otimes H\otimes J_t\otimes S \otimes\phi_{M_2}\otimes
\phi_{M_3}\;,
\end{eqnarray}
all of which are well defined and gauge invariant. The partition of nonperturbative and
perturbative contributions depends on factorization schemes. However,
a decay amplitude, as a convolution of the above factors, is
independent of factorization schemes in principle.

\section{DATABASE FOR GLOBAL FIT}

\subsection{Lightcone Distribution Amplitudes}
\label{btopp}

\begin{figure}[htbp]
\begin{minipage}[t]{0.45\linewidth}
  \centering
  \includegraphics[width=0.8\columnwidth]{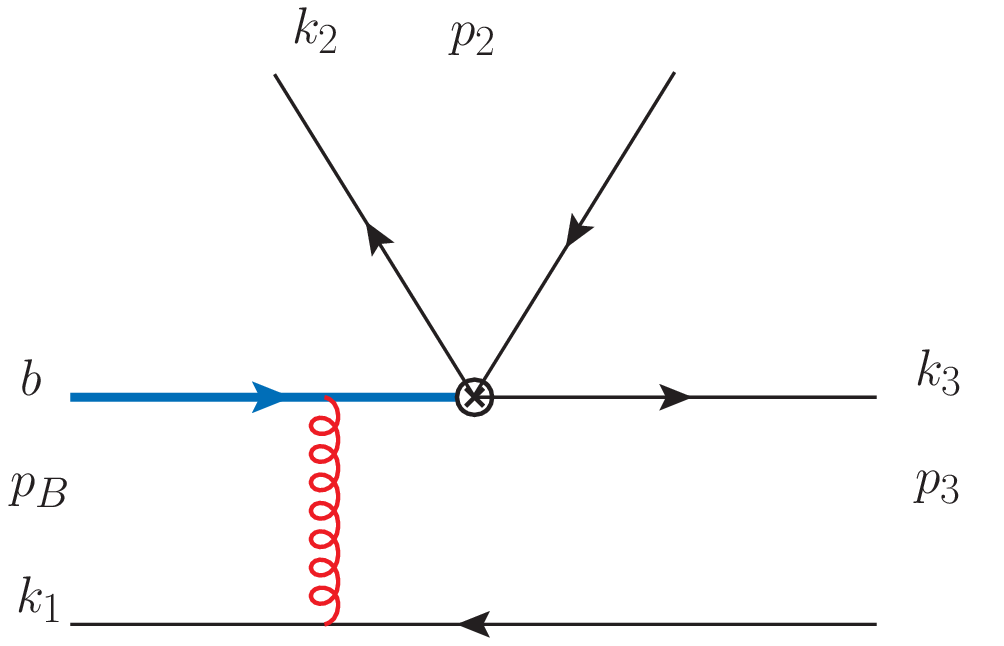}
\end{minipage}
  \caption{A LO diagram for the $\overline B(p_B)\to M_2(p_2)M_3(p_3)$ decay.}\label{fig:momentumfractions}
\end{figure}

The momenta $p_B$, $p_2$ and $p_3$ of the $\overline B$ meson, emitted meson $M_2$, and recoiling
meson $M_3$, respectively, and their associated parton momenta are chosen, in the light-cone coordinates, as
\begin{eqnarray}
	p_B = \frac{m_B}{\sqrt{2}}(1, 1, 0_T), \ \ \ \ \ \ k_1 = (x_1 \frac{m_B}{\sqrt{2}}, 0, k_{1T}), \nonumber \\
	p_2 =  \frac{m_B}{\sqrt{2}}(1, 0, 0_T), \ \ \ \ \ \ k_2 = (x_2 \frac{m_B}{\sqrt{2}}, 0, k_{2T}), \nonumber \\
	p_3 =  \frac{m_B}{\sqrt{2}}(0, 1, 0_T), \ \ \ \ \ \ k_3 = (0,x_3 \frac{m_B}{\sqrt{2}}, k_{3T}), \label{eq:momentum}
\end{eqnarray}
which are labelled in Fig.~\ref{fig:momentumfractions} with $m_B$ being the $B$ meson mass
and $x_i$ being the momentum fractions.
The light meson LCDAs are defined through the matrix elements
\begin{eqnarray}
\langle P(p)|q_{1\alpha}(0) {\bar q}_{2\beta}(z)|0\rangle
&=&\frac{i}{\sqrt{2N_c}} \int_0^1dx e^{ixp\cdot z}\left[\gamma_5\not \!
p\phi^A(x) + \gamma_5m_0\phi^P(x) +m_0\gamma_5(\not \!v\not \! n-1)
\phi^T(x)\right]_{\alpha\beta},\nonumber\\
\langle V(p,\epsilon^*_L)| q_{1\alpha} (0)\bar
q_{2\beta}(z)|0\rangle &=&\frac{-1}{\sqrt{2N_c}}\int_0^1 dx
e^{ixp\cdot z} \left[m_V\not\! \epsilon^*_L \phi_V(x) +\not\!
\epsilon^*_L\not\! p \phi_{V}^{t}(x) +m_V
\phi_V^s(x)\right]_{\alpha\beta},
\end{eqnarray}
where $N_c=3$ is the number of colors, $m_0$ is the chiral enhancement scale for the
pseudoscalar meson $P$, the dimensionless vector $v=\sqrt{2}p/M_B$ lies in the direction of the
meson momentum $p$, the dimensionless vector $n$ lies in the direction of the quark coordinate $z$
with $n\cdot v=1$, and $m_V$ ($\epsilon_L$) is the mass (longitudinal polarization vector) of the
vector meson $V$.
The light meson LCDAs are expanded as
\begin{eqnarray}
 \phi_P(x) &=&\frac{f_P}{2\sqrt{2N_c}}6x(1-x) \left[1+  a_1^f C_1^{3/2}(1-2x) +  a^f_2 C_2^{3/2}(1-2x) +  a^f_4 C_4^{3/2}(1-2x)\right],\nonumber\\
 \phi_P^P(x)&=&\frac{f_P}{2\sqrt{2N_c}}\left[1+a^f_{P2}C_2^{1/2}(1-2x)+a^f_{P4}C_4^{1/2}(1-2x)\right],\nonumber\\
 \phi_P^T(x)&=&-\frac{f_P}{2\sqrt{2N_c}}\left[C_1^{1/2}(1-2x)+a^f_{T2}C_3^{1/2}(1-2x)\right],\nonumber\\
 \phi_V(x)&=&\frac{f_V}{2\sqrt{2N_c}} 6x (1-x)\left[1+
a_{1}^{f||}C_1^{3/2} (1-2x) +a_{2}^{f||}C_2^{3/2} (1-2x) \right]\;,\nonumber\\
\phi^t_V(x) &=& \frac{3f^T_V}{2\sqrt{2N_c}}(1-2x)^2,
  \hspace*{0.5cm} \phi^s_V(x)=\frac{3f_V^T}{2\sqrt{2N_c}} (1-2x),
 \label{eq:wavefunction}
\end{eqnarray}
in terms of the orthogonal Gegenbauer polynomials
\begin{eqnarray}
 C^{1/2}_{1}(t)&=&t,\;\;\;\;C^{1/2}_{2}(t)=\frac{1}{2}(3t^2-1),\;\;\;C^{1/2}_{3}(t)=\frac{1}{2}t(5t^2-3),\nonumber\\
 C^{3/2}_{1}(t)&=&3t,\;\;\; C_2^{3/2} (t)=\frac{3}{2}(5t^2-1),\nonumber\\
 C_4^{3/2} (t)&=&\frac{15}{8}(1-14t^2+21t^4),
\end{eqnarray}
where the decay constants $f_P$, $f_V$ and $f_V^T$ 
can be extracted from leptonic decay widths such as $\Gamma(\pi \to \mu\nu)$ and $\Gamma(\tau \to \rho \nu$),
and the superscripts $f$ of the Gegenbauer moments label the species of mesons.

The $B$ meson DA is defined via the matrix element
\begin{eqnarray}
\int\frac{d^4z}{(2\pi)^4}e^{ik\cdot z}\langle 0|b_{\alpha}(0) {\bar q}_{\beta}(z)|{\overline B}(p_B)\rangle
&=&\frac{i}{\sqrt{2N_c}}\Big\{( \not\!p_B+m_B)\gamma_5\left[\phi_B(k) - \frac{\not\!n_+-\not\!n_-}{\sqrt{2}}\bar{\phi}_B(k)\right]\Big\}_{\alpha\beta},
\end{eqnarray}
with the light spectator momentum $k$, and the dimensionless vectors $n_+=(1,0,{\bf 0}_T)$ and
$n_-=(0,1,{\bf 0}_T)$. In this work we adopt the model for the $B_{(s)}$ meson DA,
\begin{eqnarray}
	\phi_{B_{(s)}}(x,b) = N_{B_{(s)}}x^2(1-x)^2 {\rm exp} \left[-\frac{m^2_{B_{(s)}}x^2}{2\omega_{B_{(s)}}^2}
	 -\frac{1}{2}\omega^2_{B_{(s)}} b^2 \right],
	\label{eq:phiB}
\end{eqnarray}
where the constant $N_{B_{(s)}}$ is fixed by the normalization condition
$\int \phi_{B_{(s)}}(x,b=0)dx=f_{B_{(s)}}/(2\sqrt{2N_c})$ with the $B_{(s)}$ meson decay constant
$f_{B_{(s)}}$, the shape parameter $\omega_{B_{(s)}}$ will
be determined in the next section, and $b$ is the impact parameter conjugate to the transverse momentum $k_T$.
It has been argued~\cite{Kurimoto:2001zj} that the contribution from $\bar{\phi}_B(k)$ is power
suppressed, so it will be neglected in the numerical analysis below.

\subsection{SU(3) Flavor Structure}

To calculate hadronic two-body  $B$ meson decay amplitudes systematically,
we introduce the following SU(3) matrix elements for various species of mesons,
\begin{eqnarray}\label{m1}
&&B^-=(1,0,0),\quad\overline{B}=(0,1,0),\quad\overline{B}^0_s=(0,0,1),\\\nonumber
&&M_{\pi^+}=M_{\rho^+}=\begin{pmatrix}
                        0 & 0 & 0 \\
                        1 & 0 & 0\\
                        0 & 0 & 0
                     \end{pmatrix},
\quad M_{K^+}=M_{K^{*+}}=\begin{pmatrix}
                        0 & 0 & 0 \\
                        0 & 0 & 0\\
                        1 & 0 & 0
                     \end{pmatrix},
\quad M_{K^0}=M_{K^{*0}}=\begin{pmatrix}
                        0 & 0 & 0 \\
                        0 & 0 & 0\\
                        0 & 1 & 0
                     \end{pmatrix},\\\nonumber
&&\sqrt{2}M_{\pi^0}=\sqrt{2}M_{\rho^0}=\begin{pmatrix}
                        1 & 0 & 0 \\
                        0 & -1 & 0\\
                        0 & 0 & 0
                     \end{pmatrix},
\quad \sqrt{2}M_{\eta_q}=\sqrt{2}M_{\omega}=\begin{pmatrix}
                        1 & 0 & 0 \\
                        0 & 1 & 0\\
                        0 & 0 & 0
                     \end{pmatrix},
\quad M_{\eta_s}=M_{\phi}=\begin{pmatrix}
                        0 & 0 & 0 \\
                        0 & 0 & 0\\
                        0 & 0 & 1
                     \end{pmatrix},\\\nonumber
&&M_{\pi^-}=M_{\rho^-}=M_{\pi^+}^T,\quad M_{K^-}=M_{K^{*-}}=M_{K^+}^T,\quad
M_{\bar K^0}=M_{\bar K^{*0}}=M_{K^0}^T,
\end{eqnarray}
which reflect  the internal structure of the flavor SU(3) group. The isosinglet mesons like
$\eta_q, \eta_s, \omega$, and $\phi$ will not be considered in the global analysis below,
but their properties are listed here for completeness.
We will take into account these hadrons, as extending the database in the future.
The matrices relevant for the heavy-to-light transitions are given by
\begin{eqnarray}
&&\delta_u=\begin{pmatrix}
                        1 & 0 & 0 \\
                        0 & 0 & 0\\
                        0 & 0 & 0
                     \end{pmatrix},
\quad \Lambda_d=\begin{pmatrix}
                        0  \\
                        1 \\
                        0
                     \end{pmatrix},
\quad \Lambda_s=\begin{pmatrix}
                        0  \\
                        0 \\
                        1
                     \end{pmatrix},
               \quad e_Q=\begin{pmatrix}
                        1 & 0 & 0 \\
                        0 & -\frac{1}{2} & 0\\
                        0 & 0 & -\frac{1}{2}
                     \end{pmatrix}.\label{m2}
\end{eqnarray}

\begin{figure}
\begin{center}
\psfig{file=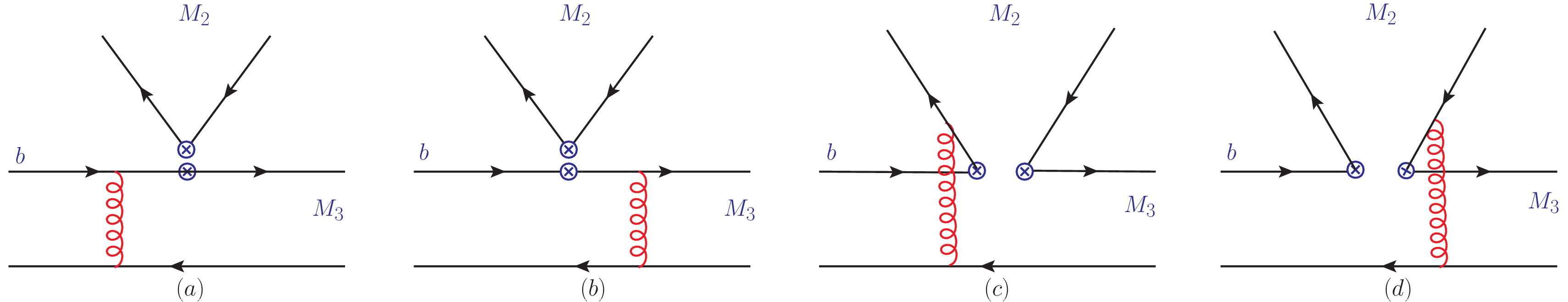,width=16.0cm,angle=0}
\end{center}
\caption{Emission diagrams with possible four-quark operator insertions. }\label{emission}
\end{figure}

\begin{figure}
\begin{center}
\psfig{file=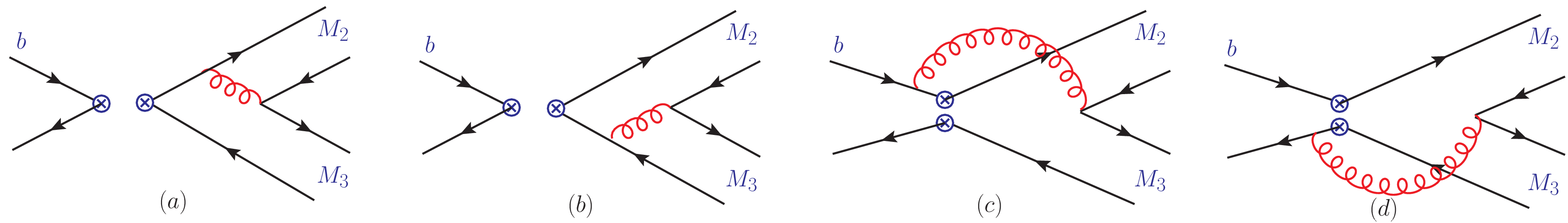,width=16.0cm,angle=0}
\end{center}
\caption{Annihilation diagrams with possible four-quark operator insertions.}
\label{annihilation}
\end{figure}

The factorization formula for a $B \to PP$ decay amplitude in Eq.~(\ref{eq:six}) can be divided
into four pieces, $F_e$ from the factorizable emission diagrams in Figs.~\ref{emission}(a) and
\ref{emission}(b), $M_e$ from the non-factorizable emission diagrams in Figs.~\ref{emission}(c) and
\ref{emission}(d), $F_a$ from the factorizable annihilation diagrams in Figs.~\ref{annihilation}(a) and
\ref{annihilation}(b), and $M_a$ from the non-factorizable annihilation diagrams in Figs.~\ref{annihilation}(c)
and \ref{annihilation}(d), each of which contains at least one hard gluon exchange.
All the diagrams receive contributions from the $(V-A)(V-A)$ operators denoted
by $LL$, from the $(V-A)(V+A)$ operators denoted by $LR$, and from the $(S-P)(S+P)$ operators denoted by $SP$.
The $(S-P)(S+P)$ operators appear under the Fierz transformation of the $(V-A)(V+A)$ ones.
The explicit expressions for the above contributions, together with
the Sudakov factors and hard kernels, are presented in Appendix~\ref{b-pp}.


We decompose the total $B \to M_2 M_3$ decay amplitude into the combination
\begin{eqnarray}
M&=&\frac{G_F}{\sqrt{2}}V_{ub}V^*_{uq}\Big[\mathcal A_u(B\to M_2 M_3)\Big]
-\frac{G_F}{\sqrt{2}}V_{tb}V^*_{tq}\Big[\mathcal A_t(B\to M_2 M_3)\Big],
\end{eqnarray}
where $A_u(B \to M_2 M_3)$ denotes the tree contribution with the product
$V_{ub}V^*_{uf}$ of the CKM matrix elements, and $A_t(B \to M_2 M_3)$ denotes the penguin contribution with
the product $V_{tb}V^*_{tf}$. These amplitudes are written, in terms of the matrices in Eqs.~(\ref{m1}) and (\ref{m2}), as
\begin{eqnarray}\label{eq:faceq}
A_u(B\to M_2 M_3)&=&\Big[F^{LL}_e( a_1)+M^{LL}_e(C_1)\Big]BM_3\delta_uM_2\Lambda_f+\Big[F^{LL}_e(a_2)+M^{LL}_e(C_2)\Big]BM_3\Lambda_fTr[\delta_uM_2]\\\nonumber
&+&\Big[F^{LL}_{ann}(a_1)+M^{LL}_{ann}(C_1)\Big]B\delta_uM_3M_2\Lambda_f
+\Big[F^{LL}_{ann}(a_2)+M^{LL}_{ann}(C_2)\Big]B\Lambda_fTr[\delta_uM_3M_2],\\\nonumber
A_t(B\to M_2 M_3)&=&\Big[F^{LL}_e(a_3)+F^{LR}_e(a_5)+M^{LL}_e(C_4)+M^{SP}_e(C_6)\Big]BM_3\Lambda_fTr[M_2]\\\nonumber
&+&\Big[F^{LL}_e(a_4)+F^{SP}_e(a_6)+M^{LL}_e(C_3)+M^{LR}_e(C_5)\Big]BM_3M_2\Lambda_f\\\nonumber
&+&\Big[F^{LR}_e(a_7)+F^{LL}_e(a_9)+M^{SP}_e(C_8)+M^{LL}_e(C_{10})\Big]BM_3\Lambda_fTr[e_QM_2]\\\nonumber
&+&\Big[F^{SP}_e(a_8)+F^{LL}_e(a_{10})+M^{LR}_e(C_7)+M^{LL}_e(C_{9})\Big]BM_3e_QM_2\Lambda_f\\\nonumber
&+&\Big[F^{LL}_{ann}(a_3)+F^{LR}_{ann}(a_5)+M^{LL}_{ann}(C_4)+M^{SP}_{ann}(C_6)\Big]B\Lambda_fTr[M_3M_2]\\\nonumber
&+&\Big[F^{LL}_{ann}(a_4)+F^{SP}_{ann}(a_6)+M^{LL}_{ann}(C_3)+M^{LR}_{ann}(C_5)\Big]BM_3M_2\Lambda_f\\\nonumber
&+&\Big[F^{LR}_{ann}(a_7)+F^{LL}_{ann}(a_9)+M^{SP}_{ann}(C_8)+M^{LL}_{ann}(C_{10})\Big]B\Lambda_fTr[e_QM_3M_2]\\\nonumber
&+&\Big[F^{SP}_{ann}(a_8)+F^{LL}_{ann}(a_{10})+M^{LR}_{ann}(C_7)+M^{LL}_{ann}(C_{9})\Big]Be_QM_3M_2\Lambda_f,
\end{eqnarray}
with the Wilson coefficients $a_1=C_2+C_1/3$, $a_2=C_1 + C_2/3$,
$a_{2n-1}=C_{2n-1}+C_{2n}/3$, and $a_{2n} = C_{2n}+C_{2n-1}/3$ ($n \geq 2$).
The unitarity of the CKM matrix is assumed in this work. The weak phase $\phi_3(\gamma)$ is
defined via the CKM matrix element $V_{ub} \equiv |V_{ub}|e^{-i\gamma}$.

The $B \to VP$ decay amplitudes can be simply inferred from the $B\to PP$ amplitudes through the
replacements of light meson LCDAs and of a chiral enhancement scale by a vector meson mass.
For the $B\to V_2P_3$ emission and $B\to P_2V_3$ annihilation,
we apply the rule $\phi_3^{(2),(3)} \to -\phi_3^{(2),(3)}$, and further flip the signs
of the $LR$ and $SP$ amplitudes, where the subscript $(2)$ means twist 2, and $(3)$ means twist 3.
For the $B\to V_2P_3$ annihilation, we apply $\phi_3^{(2)} \to -\phi_3^{(2)}$
and $\phi_3^{(3)} \to \phi_3^{(3)}$, and further flip the signs of the $LR$ and $SP$ amplitudes.
The above rule holds for both the pseudoscalar $P$ and vector $V$ mesons, and for the factorizable
and nonfactorizable diagrams. For the $B\to P_2V_3$ emission, we apply
$\phi_3^{(2)} \to -\phi_3^{(2)}$ and $\phi_3^{(3)} \to \phi_3^{(3)}$, and further flip the signs
of the nonfactorizable $LR$ amplitudes and the factorizable $SP$ amplitudes.

As shown in Eq.~(\ref{eq:wavefunction}), there are 9 Gegenbauer moments $a^f$'s in total for the twist-2
pseudoscalar LCDA $\phi_P(x)$ and twist-3 LCDAs $\phi_P^P(x)$ and $\phi_P^T(x)$.
Note that the Gegenbauer moment $a_1^\pi$ vanishes due to the isospin symmetry, and $a_{P4}^f$
are not included in the fit, because they cannot be constrained effectively under the current
limited experimental accuracy.
Thus, a $B \to PP$ decay amplitude contains $9\times 9$ combinations of the Gegenbauer moments,
 \begin{eqnarray}
 M&\sim & \sum_{n,m=1}^{9}a^f_n a^f_m M_{nm}, \label{mnm}
\end{eqnarray}
where the product of the Gegenbauer moments $a^f_na^f_m$ has been factored out explicitly. We
compute the factorization formula $M_{nm}$, which involves only the Gegenbauer polynomials
associated with $a^f_n$ and $a^f_m$, to establish a $9\times 9$ database.
Each database has 20 sets of values, corresponding to the
Wilson coefficients $a_1 \cdots a_{10}$ and $C_1 \cdots C_{10}$ in Eq.~(\ref{eq:faceq}).
To analyze the $B \to VP$ decays, we construct a $9\times 4\times 2$ database for $M_{nm}$
in a similar manner, where $\times 2$ is attributed to the two possible final states $P_2V_3$ and $V_2P_3$.
The inputs for the Fermi constant, the meson decay constants, the meson masses, and
the chiral enhancement scale are the same as in~\cite{Ali:2007ff}, and the magnitudes of the CKM
matrix elements are referred to~\cite{Zyla:2020zbs} in the above computations.

\section{NUMERICAL RESULTS}
\subsection{Least-Square Fit and Bayesian Analysis}

We determine the Gegenbauer moments and the weak phase $\phi_3(\gamma)$ by fitting the branching
ratios and direct CP asymmetries formulated from the decay amplitudes in Eq.~(\ref{mnm}) to experimental data
using the nonlinear least-$\chi^2$ (lsq) method \cite{Peter:2020}. The lsq method minimizes the summed residual $S$,
\begin{eqnarray}\label{eq:lsq_chi2}
S= \sum_{i=1}^{n} r_i^2,\;\;\;\;r_i = y_i - \hat{y_i}, 
\end{eqnarray}
where $r_i$ is the residual at the $i$-th point $x_i$, $y_i$ ($\hat{y_i}$) represents the experimental data
(response value), and $n$ is the number of data points. One defines a model function
$\hat{y_i} = f(x_i, \vec{\beta})$,
where the vector $\vec{\beta}$ contains the $m$ adjustable parameters considered in the fit.
The minimum of Eq.~(\ref{eq:lsq_chi2}) is obtained by equating the gradient to zero,
\begin{eqnarray}\label{eq:lsq_gra}
	\frac{\partial S}{\partial \beta_j} = 2\sum_i r_i\frac{\partial f(x_i, \vec{\beta})}{\partial \beta_j} = 0,  \;\;\;\; j=1,2,...,m.
\end{eqnarray}
For a linear model, $f(x_i, \vec{\beta})$ can be decomposed into a sum of multiple linear functions $G_j(x_i)$
with the corresponding coefficients $\beta_j$, $f(x_i, \vec{\beta}) = \sum_j \beta_j G_j(x_i)$.
One then regards the function $G_j(x_i)$ as a matrix, and solves for $\beta_j$ from Eq.~(\ref{eq:lsq_gra}) with
the data $y_i$.

A non-linear model is more subtle, to which there is no closed-form solution in general.
A possible approach is to select some initial values of the parameters, and refine the parameters
by iteration. At each iteration, the model function $f(x_i,\vec{\beta})$ is linearized through the
first-order Taylor series expansion at $\beta_j^k$,
\begin{eqnarray} \label{eq:non_f}
	f(x_i,\vec{\beta}) &\approx& f(x_i, \vec{\beta}^k) + \sum_j \frac{\partial f(x_i, \vec{\beta})}{\partial \beta_j} \Big(\beta_j - \beta^k_j  \Big) \\ \nonumber
		&\equiv& f(x_i,  \vec{\beta}^k)  + \sum_j J_{ij}\Delta \beta_j,
\end{eqnarray}
with $k$ being an iteration number, and $J$ being a Jacobian function.  The minimum of the residual
at this iteration is achieved by equating the gradient to zero,
\begin{eqnarray}
	\frac{\partial S}{\partial \Delta \beta_j} = -2 \sum^n_{i=1}J_{ij} \Bigg(y_i - f(x_i, \vec{\beta}^k) - \sum^m_{k=1}J_{ik} \Delta \beta_k  \Bigg) = 0,
\end{eqnarray}
where $\Delta \beta_j$ can be solved by inverting the Jacobian matrix. The parameters then take
the values $\beta^{k+1}_j = \beta^k_j + \Delta \beta_j$ for the next iteration.

Because of the approximations in the Taylor expansion and in the matrix inversion
(if the Jacobian matrix is not a square one),
no algorithm works for all nonlinear models, and fit results may be sensitive to initial conditions.
To stabilize a complicated nonlinear fit, one can perform a Bayesian analysis (conditionally biased fit).
Instead of the single $\chi^2$ term corresponding to Eq.~(\ref{eq:lsq_chi2}),
\begin{eqnarray} \label{eq:chi}
	\chi^2= \sum_{i=1}^{n}  \Big(\frac {y_i - \hat{y_i}}{\delta y_i}\Big)^2,
\end{eqnarray}
with $\delta y_i$ being the errors of experimental data, we employ a modified version
\begin{eqnarray} \label{eq:modchi}
	\chi^2_m = \chi^2  +  \chi^2_{prior}, \;\;\; \chi^2_{prior} =
	\sum_{j} \frac{(\beta_j - \tilde{\beta}_j)^2}{\tilde{\sigma}^2_{j}}.
\end{eqnarray}
The second term $\chi^2_{prior}$ is a stabilizing function, where $\tilde{\beta}_j$
are artificially introduced default values for the fitted parameters with some errors
$\tilde{\sigma}_{j}$. When the summed residual in Eq.~(\ref{eq:modchi}) are minimized,
the $\chi^2_{prior}$ term favors $\beta_j$ in the range
$(\tilde{\beta}_j - \tilde{\sigma}_{j},\tilde{\beta}_j + \tilde{\sigma}_{j})$.
The values of $\tilde{\beta}_j$ and $\tilde{\sigma}_{j}$
should be chosen reasonably according to prior physical knowledge on the fitted parameters.
In the present study we select the Gegenbauer moments derived in QCD sum rules as our
Bayesian data $\tilde{\beta}$, and five times of the errors from QCD sum rules for our
$\tilde{\sigma}$, which reduce the weight of $\chi^2_{prior}$ in the fit.

\subsection{$B_s$ Meson Distribution Amplitude}

\begin{figure}[htbp!]
   \begin{minipage}[t]{0.46\linewidth}
  \centering
  \includegraphics[width=1.0\columnwidth]{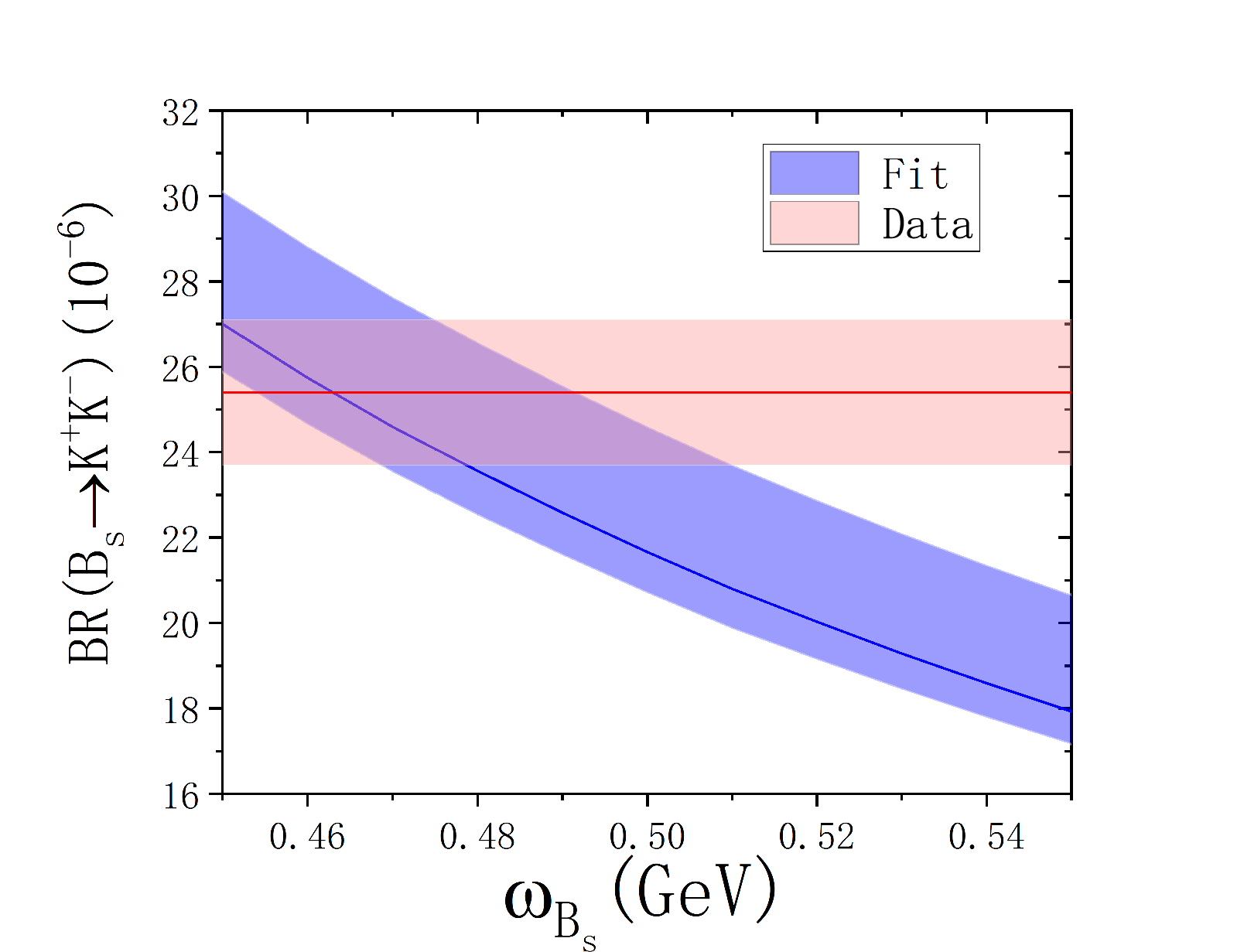}
    \end{minipage}
   \begin{minipage}[t]{0.46\linewidth}
  \centering
  \includegraphics[width=1.0\columnwidth]{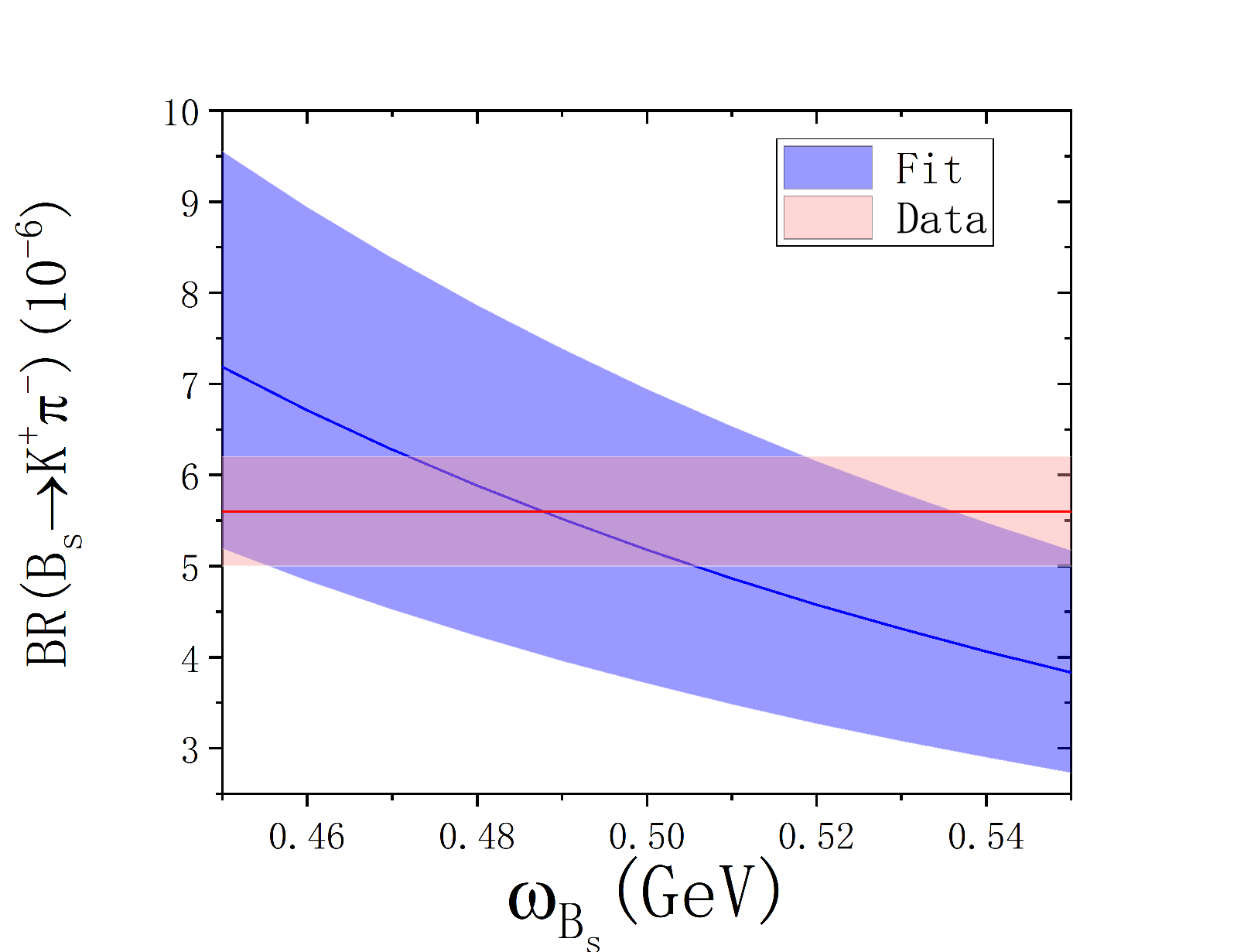}
  \end{minipage}
  \begin{minipage}[t]{0.46\linewidth}
  \centering
  \includegraphics[width=1.0\columnwidth]{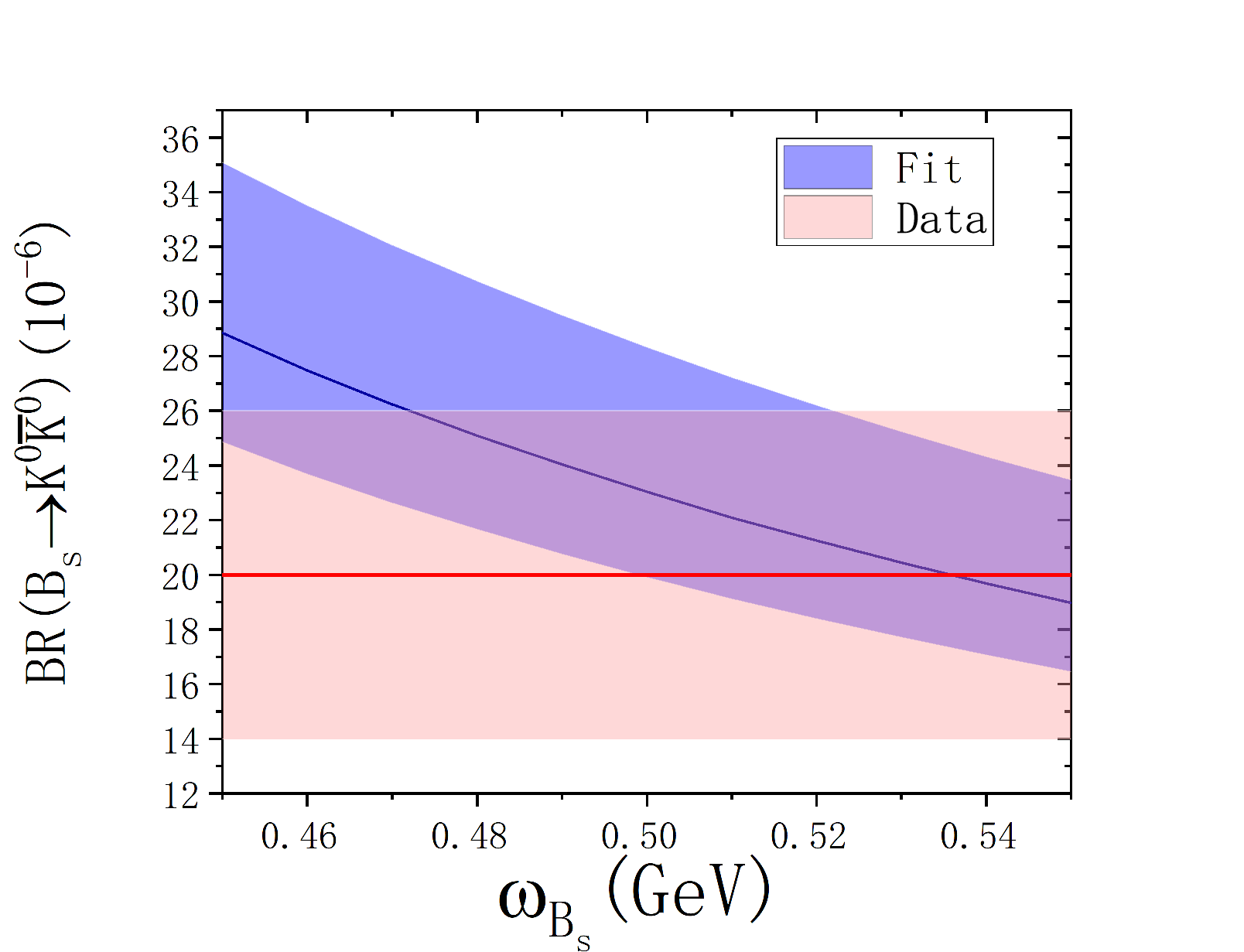}
    \end{minipage}
   \begin{minipage}[t]{0.46\linewidth}
  \centering
  \includegraphics[width=1.0\columnwidth]{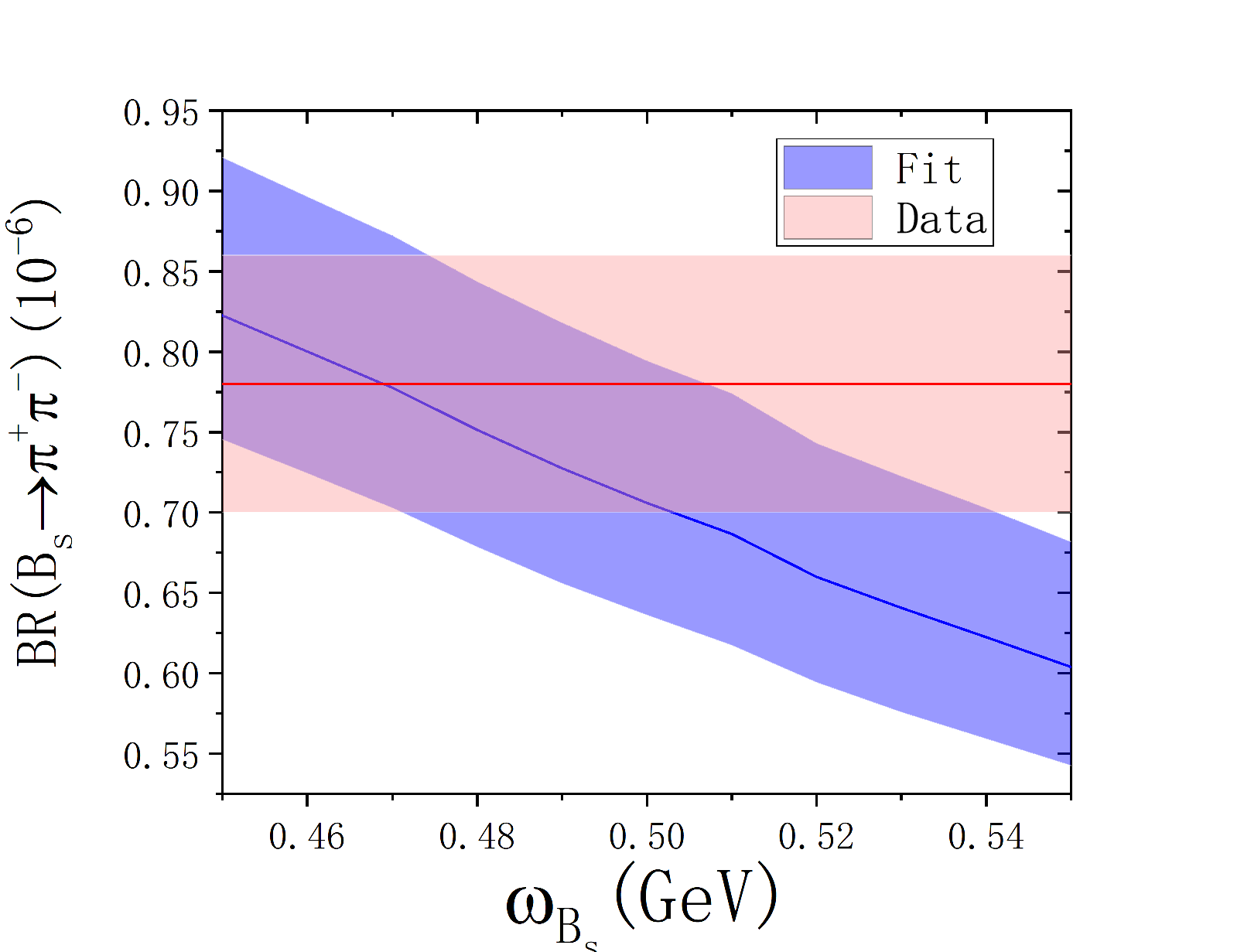}
  \end{minipage}
\caption{Dependencies of the experimental data and the reconstructed data on $\omega_{B_s}$.}\label{omegab}
\end{figure}

We point out that the $B_{(s)}$ meson DA appears in all the $B_{(s)}\to M_2M_3$
decay amplitudes, so it is difficult to extract the shape parameters $\omega_{B_{(s)}}$
in a global fit. The investigation in \cite{Kurimoto:2001zj} shows that $\omega_B=0.4$ GeV
for the $B$ meson DA leads to reasonable results for the $B\to\pi$ transition form factors, which
agree with those from light-cone sum rules and lattice QCD. Therefore, we choose this value as
the input, and perform the global fit to determine the Gegenbauer moments of the light meson LCDAs.
As to the shape parameter $\omega_{B_s}$ in the $B_s$ meson DA, it is observed that the Gegenbauer
moments fitted from the $B_{s}\to PP$ data are not sensitive to its variation: we scan the range
of $\omega_{B_s}$ from 0.4 GeV to 0.6 GeV, and make sure that the fitted Gegenbauer moments are relatively
stable in the range 0.45 GeV $<\omega_{B_s}<$ 0.55 GeV. We then select the data of several precisely
measured channels, $B_s \to K^+ K^-$, $ K^+ \pi^-$, $K^0 \bar K^0$, and
$\pi^+\pi^-$, and compare them with the reconstructed data from the PQCD factorization formulas.
The comparison displayed in Fig.~\ref{omegab}, where the red bands are from the experimental data and
the blue bands are from the reconstructed data, indicates that the choice $\omega_{B_s}=0.48$ GeV is
preferred: with this value of $\omega_{B_s}$, the PQCD results from the fit accommodate the selected
four piece of data simultaneously.

\subsection{Global Fit}
We fit the PQCD factorization formulas with the database constructed in the
previous section to the measured branching ratios and direct CP asymmetries $A_{CP}$ in the
$B_{(s)} \to PP, VP$ decays, which are collected in the left columns of
Table~\ref{tab:fitproces}. The $A_{CP}$ data marked in red, which have larger errors,
do not provide a strong constraint in the fit. Note that the LHCb Collaboration has updated their
measurement of $A_{CP}$ in the $B^0\to \bar{K}^0\pi^0$ mode, which reads $-13.8\pm 2.5$ \cite{Aaij:2020wnj}.
The data of those modes, which are greatly affected by subleading contributions according to the
existent PQCD calculations \cite{Li:2006jv,Rui:2011dr,Ali:2007ff}, namely, suffer significant theoretical uncertainties,
are excluded in our fit. The Gegenbauer moments of both the twist-2 and twist-3 LCDAs from a joint fit,
corresponding to the shape parameters $w_B=0.4$ GeV and $w_{B_s}=0.48$ GeV, are listed in Tables~\ref{tab:joint}
with $\chi^2 / d.o.f.=0.77$. The errors in our fit mainly arise from the experimental uncertainty.
The $\chi^2_{prior}$ term in the Bayesian analysis introduces little error to the fit results.
Some higher-order moments or moments of higher-twist LCDAs cannot be constrained effectively due to the
current limited experimental accuracy. This is the reason why the values of the moments
$a_{P2}^K$, $a_{T2}^K$ and $a_{1}^{K*\parallel}$ are not presented in Table~\ref{tab:joint}.

It is seen that some fitted Gegenbauer moments, like $a_{2}^{\rho\parallel}$, $a_{2}^K$ and
$a_{2}^{K^*\parallel}$, agree well with those from QCD sum rules \cite{Ball:2004ye,Ball:2007rt}
within $1\sigma$ error, which are listed in Table~\ref{tab:sumrule} for comparison.
We stress that our fit is based on the LO PQCD factorization formulas, and that next-to-leading-order (NLO)
corrections change the heavy-to-light transition form factors by about $30\%$ \cite{Li:2012bp,Wang:2012ab}.
It is difficult to estimate how much systematic error is caused by NLO effects for the fitting at LO,
because NLO corrections to the non-factorizable amplitudes in hadronic two-body $B$ meson decays have
not yet been completed in the PQCD approach. Higher-power contributions (for example, the
power-suppressed contribution from another $B$ meson DA $\bar\phi_B$ in Eq.~(\ref{eq:phiB}) was
shown to be of the same order as the NLO one in \cite{Yang:2020xal}) have not been taken into account
either. Therefore, it is likely that some fitted Gegenbauer moments, such as $a^{\pi}_2$,
differ more significantly from those in QCD sum rules. The outcome of $a_{P2}^\pi$,
slightly larger than unity, can be reduced by including the higher moment $a_{P4}^\pi$ into the
fit. As explained before, $a_{P4}^\pi$ is not considered here, because it cannot be constrained effectively under
the current experimental accuracy. It is worth mentioning that
the weak phase $\phi_3(\gamma)$ is found to be   $(75.2\pm2.9)^\circ$, consistent with
the value $(72.1^{+4.1}_{-4.5})^\circ$ in Particle Data Group \cite{Zyla:2020zbs}, and
$(69.8 \pm 2.1 \pm 0.9)^\circ$ from the factorization-assisted topological diagram approach~\cite{Zhou:2019crd}.
The agreements of our results with the Gegenbauer moments from sum rules and with $\phi_3(\gamma)$
extracted in other methods support the PQCD factorization for hadronic two-body $B$ meson decays.

\begin{table}[htbp!]
	\centering
	\caption{Gegenbauer moments and the $\gamma$ angle from a joint fit for the twist-2 and twist-3 LCDAs.}
	\begin{tabular}{|l|c|c|c|c|c|c|c|c|c|c|c|c|c|c|c|}
		\hline
		 &$a_{1}^\pi$             &$a_{2}^\pi$                &$a_{4}^\pi$                &$a_{P2}^\pi$               &$a_{T2}^\pi$               &$a_{1}^{\rho\parallel}$            &$a_{2}^{\rho\parallel}$     & \\ \hline
\ fit \  &$-$                     &$0.644\pm0.075$            &$-0.41\pm0.098$            &$1.08\pm0.15$              &$-0.48\pm0.33$             &$0$                                &$0.16\pm0.084$              & \\ \hline
         &$a_{1}^K$               &$a_{2}^K$                  &$a_{4}^K$                  &$a_{P2}^K$                 &$a_{T2}^K$                 &$a_{1}^{K^*\parallel}$             &$a_{2}^{K^*\parallel}$      &$\gamma$\\ \hline
\ fit \ &$0.331\pm0.082$          &$0.28\pm0.10$              &$-0.398\pm0.073$           &$-$                        &$-$                        &$-$                                &$0.137\pm0.029$  &$(75.2\pm2.9)^\circ$ \\ \hline
\end{tabular}\label{tab:joint}
\end{table}

\begin{table}[htbp!]
	\centering
	\caption{Gegenbauer moments of the twist-2 LCDAs from QCD sum rules \cite{Ball:2004ye,Ball:2007rt}.}
	\begin{tabular}{|l|c|c|c|c|c|c|c|c|c|c|c|c|c|c|c|}
		\hline
		  &$a^{\pi}_1$     &$a_{2}^\pi$     &$a_{4}^\pi$          &$a_{1}^{\rho\parallel}$   &$a_{2}^{\rho\parallel}$ \\ \hline
  \ fit \ &$-$                   &$0.25\pm0.15$ &$-0.015\pm0.025$&$-$                                    &$0.15\pm0.07$ \\ \hline
          &$a_{1}^K$      &$a_{2}^K$       &$a_{4}^K$           &$a_{1}^{K^*\parallel}$    &$a_{2}^{K^*\parallel}$\\ \hline
  \ fit \ &$0.06\pm0.03$&$0.25\pm0.15$ &$-$                        &$0.03\pm0.02$                 &$0.11\pm0.09$ \\ \hline
\end{tabular}\label{tab:sumrule}
\end{table}

With the fitted Gegenbauer moments in Table~\ref{tab:joint}, we calculate the branching ratios and $A_{CP}$
in the LO PQCD approach, and present the results in the right columns of Table~\ref{tab:fitproces}.
It is observed that all the considered data, except the $B^-\to\pi^0K^{*-}$ branching ratio, are well reproduced.
The observables removed from the fit, {\it i.e.}, those suffering significant theoretical uncertainties,
are also predicted in the LO PQCD formalism, and
compared with the data in Table~\ref{tab:fit2}. The predicted branching ratios are very close to the values
obtained in the previous PQCD calculations, so the deviation from the data remains.
In particular, $A_{CP}$ in the $B^-\to\pi^-\rho^{0}$ mode has been predicted to be large and negative
in most QCD approaches \cite{Cheng:2020hyj,Li:2016tpn}, but its data are as small as
$0.009\pm0.019$ \cite{Zyla:2020zbs}. The inclusion of higher-order and higher-power
contributions to hadronic two-body $B$meson decays may improve the consistency. The theoretical errors given in
Tables~\ref{tab:fitproces} and \ref{tab:fit2} arise only from those of the fitted Gegenbauer moments
and $\phi_3 (\gamma)$. More precise measurements are urged,
and subleading contributions should be included into the PQCD framework
to strengthen the constraint on the Gegenbauer moments and to sharpen the confrontation
between theoretical predictions and experimental data.

\begin{table}[htbp!]
	\centering
	\caption{Experimental data for branching ratios and direct CP asymmetries $A_{CP}$ \cite{Zyla:2020zbs}, and the theoretical
	results derived from the fitted Gegenbauer moments in Table~\ref{tab:joint}. The data with precision
	less than 3$\sigma$ are marked in red.}
	\begin{tabular}{|l|c|c|c|c|c|}
		\hline
		\multirow{2}{*}{channel}              &\multicolumn{2}{c|}{data}                       &\multicolumn{2}{c|}{fit}\cr\cline{2-5}
	                                          &branching ratio         & $A_{CP}$              &branching ratio         & $A_{CP}$\cr \hline
$B^0\to \bar{K}^0K^0$                         &$1.21\pm0.16$           &\red{$-60\pm70$}       &$1.23\pm0.08$           &$0\pm0$\\ \hline
$B^0\to \bar{K}^0\pi^0$                       &$9.90\pm0.50$           &\red{$0\pm13$}         &$8.98\pm0.19$           &$-4.02\pm0.48$\\ \hline
$B^0\to K^-\pi^+$                             &$19.6\pm0.50$           &$-8.3\pm0.6$           &$20.3\pm0.36$           &$-8.34\pm0.36$\\ \hline
$B^0\to \pi^-\pi^+$                           &$5.12\pm0.19$           &$32\pm4$               &$5.24\pm0.17$           &$23.2\pm2.1$\\ \hline
$B^0\to \rho^0\bar{K}^{0}$                    &$3.40\pm1.10$           &\red{$4\pm20$}         &$3.06\pm0.37$           &$2.853\pm0.068$\\ \hline
$B^0\to \pi^0\bar{K}^{*0}$                    &$3.30\pm0.60$           &\red{$-15\pm13$}       &$1.73\pm0.10$           &$-6.02\pm0.6$\\ \hline
$B^0\to \pi^-\rho^+/\pi^+\rho^-$              &$23.0\pm2.30$           &\red{$13\pm6/-8\pm8$}  &$23.33\pm0.8$           &$-24.3\pm1/8.1\pm1.1$\\ \hline
$B^-\to K^0K^-$                               &$1.31\pm0.17$           &\red{$4\pm14$}         &$1.47\pm0.09$           &$22.5\pm2.7$\\ \hline
$B^-\to \pi^0K^-$                             &$12.9\pm0.50$           &\red{$3.7\pm2.1$}      &$12.99\pm0.23$          &$-6.44\pm0.6$\\ \hline
$B^-\to \bar{K}^0\pi^-$                       &$23.7\pm0.80$           &\red{$-1.7\pm1.6$}     &$23.15\pm0.42$          &$-2.84\pm0.24$\\ \hline
$B^-\to \rho^-\pi^0$                          &$10.9\pm1.40$           &\red{$2\pm11$}         &$8.73\pm0.25$           &$24.2\pm2.3$\\ \hline
$B^-\to \pi^0K^{*-}$                          &$6.80\pm0.90$           &\red{$-39\pm21$}       &$3.51\pm0.19$           &$-33.5\pm1.7$\\ \hline
$B^-\to K^-K^{*0}$                            &$0.59\pm0.08$           &\red{$12\pm10$}        &$0.476\pm0.022$         &$22.5\pm1.3$\\ \hline
$B_s\to K^-K^+$                               &$26.6\pm2.20$           &\red{$-14\pm11$}       &$24.8\pm1.50$           &$-8.1\pm2.3$\\ \hline
$B_s\to \pi^-\pi^+$                           &$0.7\pm0.1$             &$-$                    &$0.798\pm0.092$         &$-1.62\pm0.39$\\ \hline
$B_s\to K^0\bar{K}^{0}$                       &$20.0\pm6.00$           &$0\pm0$                &$26.2\pm1.60$           &$0\pm0$\\ \hline
$B_s\to \pi^-K^+$                             &$5.80\pm0.70$           &$22.1\pm1.5$           &$5.69\pm0.64$           &$22.1\pm1.2$\\ \hline
$B_s\to K^+K^{*-}/K^-K^{*+}$                  &$19.0\pm5.0$            &$-$                    &$15.28\pm0.90$          &$-33.8\pm1.3/53.5\pm2.4$\\ \hline
$B_s\to K^0\bar{K}^{*0}/\bar{K}^0K^{*0}$      &$20.0\pm6.00$           &$-$                    &$15.06\pm0.96$          &$0\pm0$\\ \hline
\end{tabular}
\label{tab:fitproces}
\end{table}

\begin{table}[htbp!]
\centering
\caption{LO PQCD predictions for the observables removed from the fit, and compared with those	in previous PQCD analyses \cite{Li:2006jv,Rui:2011dr,Ali:2007ff,Lu:2000hj}. }
\begin{tabular}{|l|c|c|c|c|c|c|c|c|c|c|c|c|c|c|c|}
\hline
\multirow{2}{*}{channel}&\multicolumn{2}{c|}{data}&\multicolumn{2}{c|}{fit}&PQCD \cr\cline{2-6}
	                        &branching ratio        &$A_{CP}$           &branching ratio    &$A_{CP}$           &branching ratio\cr \hline
$B^0\to K^+K^-$             &$0.078\pm0.015$        &$-$                &$0.155\pm0.027$    &$52.0\pm15.0$      & \\ \hline
$B^0\to\pi^+K^{*-}$         &$7.5\pm0.4$            &$-27\pm4$          &$4.93\pm0.28$      &$-52.0\pm2.1$      &5.1 \cite{Li:2006jv}\\ \hline
$B^0\to\pi^0\rho^0$         &$2.0\pm0.5$            &$-27\pm24$         &$0.026\pm0.0022$   &$-47\pm21$         &0.15 \cite{Rui:2011dr} \\ \hline
$B^0\to K^-\rho^+$          &$7.0\pm0.9$            &$20\pm11$          &$4.41\pm0.6$       &$48.3\pm4.9$       &4.7 \cite{Li:2006jv}\\ \hline
$B^-\to\rho^-\bar{K}^{0}$   &$7.3\pm1.2$            &$-3\pm15$          &$3.39\pm0.55$      &$3.18\pm0.55$      &3.6 \cite{Li:2006jv}\\ \hline
$B^-\to\rho^0K^{-}$         &$3.7\pm0.5$            &$37\pm1$           &$2.24\pm0.41$      &$69.7\pm3.0$       &2.5 \cite{Li:2006jv} \\ \hline
$B^-\to\pi^-\bar{K}^{*0}$   &$10.1\pm0.8$           &$-4\pm9$           &$5.17\pm0.23$      &$-0.61\pm0.19$     &5.5 \cite{Li:2006jv} \\ \hline
$B^-\to\pi^-\rho^{0}$       &$8.3\pm1.2$            &$0.009\pm0.019$    &$4.61\pm0.36$      &$-35.3\pm1.8$      &$\sim 5.39$\cite{Lu:2000hj} \\ \hline
$B_s\to\pi^-K^{*+}$         &$2.9\pm1.1$            &$-$                &$9.53\pm0.24$      &$-25.5\pm1.0  $    &7.6\cite{Ali:2007ff}\\ \hline
\end{tabular}\label{tab:fit2}
\end{table}

\subsection{Remarks and Future Developments}

A few remarks are given as follows.
\begin{itemize}
\item We have focused only on the branching ratios and direct CP asymmetries in the $B\to PP, VP$ decays,
and neglected those modes involving isosinglet mesons in the above analysis. Other observables, such as
mixing-induced CP asymmetries and polarizations in $B\to VV$ decays, can be included straightforwardly.
Though more parameters will be introduced through LCDAs for transversely polarized vector mesons, sufficient precise
measurements on polarization observables can be achieved at LHCb and Belle-II.

\item LCDAs also appear in the factorization formulas for heavy-to-light transition form factors that govern
semileptonic $B$ meson decays. One can take into account experimental constraints from these decays in the future,
in particular those from their dependence on the lepton-pair invariant mass squared $q^2$.

\item As we have pointed out, decay widths of a few modes are suppressed at LO in PQCD, and may
be well described with the inclusion of higher-order contributions~\cite{Li:2006jv,Yan:2017nlj}.
Some sources of power corrections have been explored in Refs.~\cite{Wang:2017ijn,Wang:2018wfj}.   A new database
will be established in a similar way by using the flavor structure for these radiative and power corrections,
via which the precision of a global analysis can be enhanced.

\item We did not consider all systematic and parametric  uncertainties in the current analysis, such as the ones
originating from the variations of factorization scales and nonperturbative QCD parameters.

\item If a high-precision global study  reveals notable tensions between theoretical
results and experimental data in the future, it may hint that new physics effects are inevitable.
One is then motivated to include new physics contributions, which can also be analyzed according to the
flavor structure of new physics operators.

\end{itemize}

\section{Summary}

As stated in the Introduction, nonperturbative hadron LCDAs provide a major source of theoretical uncertainties
in all the factorization-based approaches to hadronic two-body $B$ meson decays.
In this paper we have performed a global analysis of the light meson LCDAs by fitting the LO PQCD factorization formulas for
$B_{(s)}\to PP,VP$ decays to available data of branching ratios and direct CP asymmetries.
A computation code was developed based on the flavor structure of the four-quark effective
operators to establish the database, which contains the part of decay amplitudes
without the Gegenbauer moments. This database facilitates the global fit, from which the Gegenbauer
moments of the twist-2 and twist-3 LCDAs for the pseudoscalar meson $P$ ($P=\pi$, $K$) and vector meson
$V$ ($V=\rho$, $K^*$) were determined. Most of our fit results agree with the moments derived in QCD sum
rules, and those with discrepancies deserve more thorough investigation that takes into account
higher-order and higher-power corrections in the PQCD approach. The weak
phase $\phi_3(\gamma)=(75.2\pm2.9)^\circ$ in consistency with the value in Particle Data Group was also extracted.
Predictions for the modes, which were excluded in the fit due to large theoretical uncertainties,
are close to the existent PQCD results, and still deviate from the data.
To improve the consistency, subleading contributions to
hadronic two-body $B$ meson decays need to be included, when their evaluation is
completed in the future.

Since the $B_{(s)}$ meson DA appears in all the $B_{(s)}\to M_2M_3$
decay amplitudes, it is difficult to constrain the shape parameters $\omega_{B_{(s)}}$
in this DA in a global fit. The shape parameter $\omega_{B}=0.4$ GeV is an input, and
$\omega_{B_s}=0.48$ GeV is subject to a discretionary choice in the present study.
The difficulty is expected to be overcome, when data for exclusive
processes other than hadronic two-body $B$ meson decays are considered in the global fit. This is
a straightforward extension of the framework proposed here, through which the global determination
of LCDAs for other hadrons is also feasible.

\section*{Acknowledgement}
WW thanks Prof. Deshan Yang for a long discussion on a possible
global analysis under the flavor SU(3) structure in the PQCD approach.
JH thanks Dr. Dacheng Yan for useful discussions on two-body $B$ meson decays.
JH is supported by NSFC under Grant 11947215. JH, WW, and ZPX are supported in part by Natural Science
Foundation of China under Grant Nos. 11735010, U2032102, and by Natural Science Foundation of
Shanghai under Grant No. 15DZ2272100. HNL is supported by MOST of R.O.C. under Grant No.
MOST-107-2119-M-001-035-MY3. CDL is supported by National Science Foundation of China under Grant Nos. 11521505 and 12070131001.  The computing package to construct the database is available on request.

\begin{appendix}

\section{$B\to PP$ DECAY AMPLITUDES}\label{b-pp}
The explicit LO PQCD factorization formulas for the $B\to PP$ decay amplitudes from the various
current operators and topologies are presented in this appendix, with $C_F=4/3$, the Wilson coefficients
$a_i$, and $r_{i}=m_{0i}/m_B$, where $m_{0i}$ is the chiral enhancement scale:
  \begin{eqnarray}\label{eq:PPLL}
  F^{LL}_{e}(a_i)&=&8\pi
  C_Fm_{B}^4f_{P_2}\int^1_0dx_1dx_3\int^\infty_0b_1db_1b_3db_3
\phi_{B}(x_1,b_1)
  \Big\{a_i(t_a) E_e(t_a)
  \nonumber\\
  &&\times \Big[(2-x_3)\phi_3^A(x_3)+r_3(2x_3-1)(\phi_3^P(x_3)-\phi_3^T(x_3))
  \Big]h_e(x_1,1-x_3,b_1,b_3)
 \nonumber\\ && \;\;+2r_3\phi_3^P(x_3)a_i(t_a^\prime) E_e(t_a^\prime)h_e(1-x_3,x_1,b_3,b_1)
 \Big\},\label{ppefll}
  \end{eqnarray}
\begin{eqnarray}
  F^{LR}_{e}(a_i)&=&- F^{LL}_{e}(a_i),\label{ppeflr}
\end{eqnarray}
  \begin{eqnarray}
  F^{SP}_{e}(a_i)&=& 16\pi r_2
  C_Fm_B^4f_{M_2}\int^1_0dx_1dx_3\int^\infty_0b_1db_1b_3db_3
\phi_{B}(x_1,b_1)
  \Big\{a_i(t_a)E_e(t_a)\nonumber\\
  &&\times\Big[\phi_3^A(x_3)+r_3(3-x_3)\phi_3^P(x_3)+r_3(1-x_3)\phi_3^T(x_3)\Big]
  h_e(x_1,1-x_3,b_1,b_3)\nonumber\\
  &&\;\;\;+2
  r_3\phi_3^P(x_3)a_i(t^\prime_a)E_e(t_a^\prime)h_e(1-x_3,x_1,b_3,b_1)\Big\},\label{ppefsp}
  \end{eqnarray}
\begin{eqnarray}
  M^{LL}_{e}(C_i)&=&32\pi C_Fm_B^4/\sqrt{6}\int^1_0dx_1dx_2dx_3\int^\infty_0b_1db_1b_2db_2
\phi_{B}(x_1,b_1)\phi_2^A(x_2)
\nonumber\\
&&\times
\Big\{\Big[x_2\phi_3^A(x_3)+r_3(x_3-1)(\phi_3^P(x_3)+\phi_3^T(x_3))\Big]
C_i(t_b)E_e^\prime(t_b)\nonumber\\
&&~\times h_n(x_1,x_2,1-x_3,b_1,b_2)+h_n(x_1,1-x_2,1-x_3,b_1,b_2)\nonumber\\
 &&\;\;\times\Big[-(2-x_2-x_3)\phi_3^A(x_3)+r_3(1-x_3)(\phi_3^P(x_3)-\phi_3^T(x_3))\Big]
 C_i(t_b^\prime) E_e^\prime(t_b^\prime)\Big\},\label{ppenll}
 \end{eqnarray}
\begin{eqnarray}
  M^{LR}_{e}(C_i)&=&32\pi C_Fm_B^4r_2/\sqrt{6}
      \int^1_0dx_1dx_2dx_3\int^\infty_0b_1db_1b_2db_2\phi_{B}(x_1,b_1)\nonumber\\
     &&\times \Big\{h_n(x_1,x_2,1-x_3,b_1,b_2)\Big[r_3(1-x_3)(\phi_2^P(x_2)+\phi_2^T(x_2))
     (\phi_3^P(x_3)-\phi_3^T(x_3))\nonumber\\
     &&\;\;+r_3x_2(\phi_2^P(x_2)-\phi_2^T(x_2))(\phi_3^P(x_3)+\phi_3^T(x_3))
     +x_2\phi_3^A(x_3)(\phi_2^P(x_2)-\phi_2^T(x_2))\Big]C_i(t_b)
         E_e^\prime(t_b) \nonumber\\
       &&\;\;+h_n(x_1,1-x_2,1-x_3,b_1,b_2)\Big[(x_2-1)\phi_3^A(x_3)(\phi_2^P(x_2)+\phi_2^T(x_2))\nonumber\\
       &&\;\;+r_3(x_2-1)(\phi_2^P(x_2)+\phi_2^T(x_2))(\phi_3^P(x_3)+\phi_3^T(x_3))\nonumber\\
       &&\;\;+r_3(x_3-1)(\phi_2^P(x_2)
       -\phi_2^T(x_2))(\phi_3^P(x_3)-\phi_3^T(x_3))\Big]C_i(t^\prime_b)
       E_e^\prime(t_b^\prime)\Big\},\label{ppenlr}
 \end{eqnarray}
\begin{eqnarray}
 M^{SP}_{e}(C_i) &=&32\pi C_F
m_B^4/\sqrt{6}\int^1_0dx_1dx_2dx_3\int^\infty_0b_1db_1b_2db_2
\phi_{B}(x_1,b_1)\phi_2^A(x_2)
\nonumber\\
&&\times\Big\{
\Big[(x_3-1-x_2)\phi_3^A(x_3)+r_3(1-x_3)(\phi_3^P(x_3)-\phi_3^T(x_3))\Big]\nonumber\\
&&\;\;\times
C_i(t_b)E_e^\prime(t_b)h_n(x_1,x_2,1-x_3,b_1,b_2)+C_i(t_b^\prime)
E^\prime_e(t_b^\prime)\nonumber\\
&&\;\;\times
 \Big[(1-x_2)\phi_3^A(x_3)+r_3(x_3-1)(\phi_3^T(x_3)+\phi_3^P(x_3))\Big]h_n(x_1,1-x_2,1-x_3,b_1,b_2)\Big\}.
 \label{ppensp}
\end{eqnarray}
\begin{eqnarray}
  F^{LL}_{ann}(a_i)&=&8\pi
C_Fm_B^4f_{B}\int^1_0dx_2dx_3\int^\infty_0b_2db_2b_3db_3\Big\{a_i(t_c)
E_a(t_c)
\nonumber\\
&&
\times\Big[-x_3\phi_2^A(x_2)\phi_3^A(x_3)-2r_2r_3(1+x_3)\phi_2^P(x_2)\phi_3^P(x_3)\nonumber
\\
&&\;\;+2r_2r_3(1-x_3)\phi_2^P(x_2)\phi_3^T(x_3)\Big]h_a(1-x_2,x_3,b_2,b_3)\nonumber
\\
&&+\Big[(1-x_2)\phi_2^A(x_2)
\phi_3^A(x_3)+2r_2r_3(2-x_2)\phi_2^P(x_2)\phi_3^P(x_3)
 +2r_2r_3x_2\phi_3^P(x_3)\phi_2^T(x_2)\Big]\nonumber\\
 &&\times a_i(t_c^\prime)
E_a(t_c^\prime)h_a(x_3,1-x_2,b_3,b_2)\Big\}.\label{ppafll}
 \end{eqnarray}
\begin{eqnarray}
   F^{LR}_{ann}(a_i)=  F^{LL}_{ann}(a_i),\label{ppaflr}
\end{eqnarray}
 \begin{eqnarray}
  F^{SP}_{ann}(a_i)&=&16\pi
  C_Fm_B^4f_{B}\int^1_0dx_2dx_3\int^\infty_0b_2db_2b_3db_3
  \Big\{\Big[2r_2\phi_2^P(x_2)\phi_3^A(x_3)\nonumber\\
  &&\;\;+x_3r_3\phi_2^A(x_2)(\phi_3^P(x_3)
  -\phi_3^T(x_3))\Big]
 a_i(t_c) E_a(t_c)h_a(1-x_2,x_3,b_2,b_3)\nonumber\\
  &&\;\;+\Big[2r_3\phi_2^A(x_2)\phi_3^P(x_3)+r_2(1-x_2)(\phi_2^P(x_2)+\phi_2^T(x_2))\phi_3^A(x_3)
  \Big]\nonumber\\
  &&\;\;\times
  a_i(t_c^\prime)E_a(t_c^\prime)h_a(x_3,1-x_2,b_3,b_2)\Big\}.\label{ppafsp}
\end{eqnarray}
\begin{eqnarray}
 M_{ann}^{LL}(C_i)&=&32\pi C_Fm_B^4/\sqrt
 {6}\int^1_0dx_1dx_2dx_3\int^\infty_0b_1db_2b_2db_2\phi_{B}(x_1,b_1)\nonumber\\
 &&\times
 \Big\{h_{na}(x_1,1-x_2,1-x_3,b_1,b_2)\Big[(x_2-1)\phi_2^A(x_2)\phi_3^A(x_3)\nonumber\\
 &&\;\;-r_2r_3\left((x_3-1)(\phi_2^P(x_2)+\phi_2^T(x_2))(\phi_3^P(x_3)-\phi_3^T(x_3))
 +4\phi_2^P(x_2)\phi_3^P(x_3)\right.\nonumber\\
 &&\;\;\left.-x_2(\phi_2^P(x_2)-\phi_2^T(x_2))(\phi_3^P(x_3)+\phi_3^T(x_3))\right)\Big]C_i(t_d)
 E_a^\prime(t_d)\nonumber\\
 &&\;\;+h_{na}^\prime(x_1,1-x_2,1-x_3,b_1,b_2)\Big[x_3\phi_2^A(x_2)\phi_3^A(x_3)\nonumber\\
 &&\;\;+r_2r_3\left(x_3(\phi_2^P(x_2)-\phi_2^T(x_2))(\phi_3^P(x_3)+\phi_3^T(x_3))\right.\nonumber\\
 &&\;\;\left.+(1-x_2)(\phi_2^P(x_2)+\phi_2^T(x_2))(\phi_3^P(x_3)-\phi_3^T(x_3))\right)\Big]C_i(t_d^\prime)
 E_a^\prime(t_d^\prime)\Big\},\label{ppanll}
 \end{eqnarray}
 \begin{eqnarray}
 M_{ann}^{LR}(C_i)&=&32\pi C_Fm_B^4/\sqrt
 {6}\int^1_0dx_1dx_2dx_3\int^\infty b_1db_1b_2db_2\phi_{B}(x_1,b_1)\nonumber\\
 &&\;\;\times\Big\{h_{na}(x_1,1-x_2,1-x_3,b_1,b_2)\Big[r_2(1+x_2)\phi_3^A(x_3)(\phi_2^P(x_2)-\phi_2^T(x_2))\nonumber\\
 &&\;\;+r_3(x_3-2)\phi_2^A(x_2)(\phi_3^P(x_3)+\phi_3^T(x_3))\Big]C_i(t_d)E_a^\prime(t_d)
 \nonumber\\
 &&\;\;+h_{na}^\prime
 (x_1,1-x_2,1-x_3,b_1,b_2)\Big[r_2(1-x_2)\phi_3^A(x_3)(\phi_2^P(x_2)-\phi_2^T(x_2))\nonumber\\
 &&\;\;-r_3x_3\phi_2^A(x_2)(\phi_3^P(x_3)+\phi_3^T(x_3))\Big]  C_i(t_d^\prime)E_a^\prime(t_d^\prime)
 \Big\},\label{ppanlr}
 \end{eqnarray}
 \begin{eqnarray}\label{eq:PPSP}
 M_{ann}^{SP}(C_i)&=&32\pi C_F m_B^4/\sqrt {6}\int^1_0dx_1dx_2dx_3\int^\infty_0b_1db_1b_2db_2
 \phi_{B}(x_1,b_1)\nonumber\\
 &&\times \Big\{C_i(t_d)E_a^\prime(t_d)h_{na}(x_1,1-x_2,1-x_3,b_1,b_2)\Big[-x_3
 \phi_2^A(x_2)\phi_3^A(x_3)\nonumber\\
 &&\;\;-4r_2r_3\phi_2^P(x_2)\phi_3^P(x_3)+r_2r_3(1-x_3)(\phi_2^P(x_2)-\phi_2^T(x_2))
 (\phi^P_3(x_3)+\phi_3^T(x_3))\nonumber\\
 &&\;\;+r_2r_3x_2(\phi_2^P(x_2)+\phi_2^T(x_2))(\phi^P_3(x_3)-\phi_3^T(x_3))\Big]
 \nonumber\\
 &&\;\;+C_i(t_d^\prime)
 E_a^\prime(t_d^\prime)h_{na}^\prime(x_1,1-x_2,1-x_3,b_1,b_2)
  \Big[(1-x_2)\phi_2^A(x_2)\phi_3^A(x_3)\nonumber
 \\
 &&\;\;+r_2r_3(1-x_2)(\phi_2^P(x_2)-\phi_2^T(x_2))
 (\phi_3^P(x_3)+\phi_3^T(x_3))\nonumber\\
 &&\;\;+r_2r_3x_3(\phi_2^P(x_2)+\phi_2^T(x_2))(\phi_3^P(x_3)-\phi_3^T(x_3))\Big]\Big\}.
 \label{ppansp}
 \end{eqnarray}



The hard scales involved in the above decay amplitudes are defined by
\begin{eqnarray}
t_a&=&\mbox{max}\{{\sqrt
{1-x_3}m_{B},1/b_1,1/b_3}\},\nonumber\\
t_a^\prime&=&\mbox{max}\{{\sqrt
{x_1}m_{B},1/b_1,1/b_3}\},\nonumber\\
t_b&=&\mbox{max}\{\sqrt
{x_1(1-x_3)}m_{B},\sqrt{|x_2-x_1|(1-x_3)}m_{B},1/b_1,1/b_2\},\nonumber\\
t_b^\prime&=&\mbox{max}\{\sqrt{x_1(1-x_3)}m_{B},\sqrt
{|1-x_1-x_2|(1-x_3)}m_{B},1/b_1,1/b_2\},\nonumber\\
t_c&=&\mbox{max}\{\sqrt{x_3}m_{B},1/b_2,1/b_3\},\nonumber\\
t_c^\prime
&=&\mbox{max}\{\sqrt {1-x_2}m_{B},1/b_2,1/b_3\},\nonumber\\
t_d&=&\mbox{max}\{\sqrt {(1-x_2)x_3}m_{B},
\sqrt{1-(x_2-x_1)(1-x_3)}m_{B},1/b_1,1/b_2\},\nonumber\\
t_d^\prime&=&\mbox{max}\{\sqrt{x_3(1-x_2)}m_{B},\sqrt{|x_1-(1-x_2)|x_3}m_{B},1/b_1,1/b_2\}.
\end{eqnarray}
The hard kernels $h$  in the decay amplitudes consist of two parts,
the jet function $J_t(x_i)$ derived in the threshold
resummation and the Fourier transformation of the virtual particle propagators:
\begin{eqnarray}
h_e(x_1,x_3,b_1,b_3)&=&\left[\theta(b_1-b_3)I_0(\sqrt
x_3m_{B}b_3)K_0(\sqrt
x_3 m_{B}b_1)\right.\\
&& \left.+\theta(b_3-b_1)I_0(\sqrt x_3m_{B}b_1)K_0(\sqrt
x_3m_{B}b_3)\right]K_0(\sqrt {x_1x_3}m_{B}b_1)J_t(x_3),\nonumber\\
h_n(x_1,x_2,x_3,b_1,b_2)&=&\left[\theta(b_2-b_1)K_0(\sqrt
{x_1x_3}m_{B}b_2)I_0(\sqrt
{x_1x_3}m_{B}b_1)\right. \nonumber\\
&&\;\;\;\left. +\theta(b_1-b_2)K_0(\sqrt
{x_1x_3}m_{B}b_1)I_0(\sqrt{x_1x_3}m_{B}b_2)\right]\nonumber\\
&&\times
\left\{\begin{array}{ll}\frac{i\pi}{2}H_0^{(1)}(\sqrt{(x_2-x_1)x_3}
m_{B}b_2),& x_1-x_2<0\\
 K_0(\sqrt{(x_1-x_2)x_3}m_{B}b_2),& x_1-x_2>0
\end{array}
\right.
\end{eqnarray}
\begin{eqnarray}
h_a(x_2,x_3,b_2,b_3)&=&(\frac{i\pi}{2})^2
J_t(x_3)\Big[\theta(b_2-b_3)H_0^{(1)}(\sqrt{x_3}m_{B}b_2)J_0(\sqrt
{x_3}m_{B}b_3)\nonumber\\
&&\;\;+\theta(b_3-b_2)H_0^{(1)}(\sqrt {x_3}m_{B}b_3)J_0(\sqrt
{x_3}m_{B}b_2)\Big]H_0^{(1)}(\sqrt{x_2x_3}m_{B}b_2),
\nonumber \\ \\
h_{na}(x_1,x_2,x_3,b_1,b_2)&=&\frac{i\pi}{2}\left[\theta(b_1-b_2)H^{(1)}_0(\sqrt
{x_2(1-x_3)}m_{B}b_1)J_0(\sqrt {x_2(1-x_3)}m_{B}b_2)\right. \nonumber\\
&&\;\;\left.
+\theta(b_2-b_1)H^{(1)}_0(\sqrt{x_2(1-x_3)}m_{B}b_2)J_0(\sqrt
{x_2(1-x_3)}m_{B}b_1)\right]\nonumber\\
&&\;\;\;\times K_0(\sqrt{1-(1-x_1-x_2)x_3}m_{B}b_1),
\\
h_{na}^\prime(x_1,x_2,x_3,b_1,b_2)&=&\frac{i\pi}{2}\left[\theta(b_1-b_2)H^{(1)}_0(\sqrt
{x_2(1-x_3)}m_{B}b_1)J_0(\sqrt{x_2(1-x_3)}m_{B}b_2)\right. \nonumber\\
&&\;\;\;\left. +\theta(b_2-b_1)H^{(1)}_0(\sqrt
{x_2(1-x_3)}m_{B}b_2)J_0(\sqrt{x_2(1-x_3)}m_{B}b_1)\right]\nonumber\\
&&\;\;\;\times
\left\{\begin{array}{ll}\frac{i\pi}{2}H^{(1)}_0(\sqrt{(x_2-x_1)(1-x_3)}m_{B}b_1),&
x_1-x_2<0\\
K_0(\sqrt {(x_1-x_2)(1-x_3)}m_{B}b_1),& x_1-x_2>0\end{array}\right.
,
\end{eqnarray}
with the Bessel function $H_0^{(1)}(z) = \mathrm{J}_0(z) + i\, \mathrm{Y}_0(z)$.
The following approximate parametrization for the jet function has been proposed for
convenience \cite{Kurimoto:2001zj},
\begin{eqnarray}
J_t(x)=\frac{2^{1+2c}\Gamma(3/2+c)}{\sqrt{\pi}\Gamma(1+c)} [x(1-x)]^c,
\label{eq:trs}
\end{eqnarray}
with the parameter $c\approx 0.3$. The prefactor in the above expression is chosen
to obey the normalization $\int_0^1 J_t(x)dx=1$.
The jet function $J_t(x)$ gives a
very small numerical effect to the nonfactorizable amplitude~\cite{Li:2001ay},
so it is dropped from $h_n$ and $h_{na}$.


The evolution factors $E^{(\prime)}_e$ and $E^{(\prime)}_a$ are written as
\begin{eqnarray}
E_e(t)&=&\alpha_s(t) \exp[-S_B(t)-S_3(t)],
 \ \ \ \
 E'_e(t)=\alpha_s(t)
 \exp[-S_B(t)-S_2(t)-S_3(t)]|_{b_1=b_3},\\
E_a(t)&=&\alpha_s(t)
 \exp[-S_2(t)-S_3(t)],\
 \ \ \
E'_a(t)=\alpha_s(t) \exp[-S_B(t)-S_2(t)-S_3(t)]|_{b_2=b_3},
\end{eqnarray}
in which the Sudakov exponents are given by
\begin{eqnarray}
S_B(t)&=&s\left(x_1\frac{m_B}{\sqrt
2},b_1\right)+\frac{5}{3}\int^t_{1/b_1}\frac{d\bar \mu}{\bar
\mu}\gamma_q(\alpha_s(\bar \mu)),\\
S_2(t)&=&s\left(x_2\frac{m_B}{\sqrt
2},b_2\right)+s\left((1-x_2)\frac{m_B}{\sqrt
2},b_2\right)+2\int^t_{1/b_2}\frac{d\bar \mu}{\bar
\mu}\gamma_q(\alpha_s(\bar \mu)),
\end{eqnarray}
with the quark anomalous dimension $\gamma_q=-\alpha_s/\pi$. Replacing the
kinematic variables of $M_2$ by those of $M_3$ in $S_2$, we get the
expression for $S_3$. The function $s(Q,b)$ is expressed as
\begin{eqnarray}
s(Q,b)&=&~~\frac{A^{(1)}}{2\beta_{1}}\hat{q}\ln\left(\frac{\hat{q}}
{\hat{b}}\right)-
\frac{A^{(1)}}{2\beta_{1}}\left(\hat{q}-\hat{b}\right)+
\frac{A^{(2)}}{4\beta_{1}^{2}}\left(\frac{\hat{q}}{\hat{b}}-1\right)
\nonumber \\
&&-\left[\frac{A^{(2)}}{4\beta_{1}^{2}}-\frac{A^{(1)}}{4\beta_{1}}
\ln\left(\frac{e^{2\gamma_E-1}}{2}\right)\right]
\ln\left(\frac{\hat{q}}{\hat{b}}\right)
\nonumber \\
&&+\frac{A^{(1)}\beta_{2}}{4\beta_{1}^{3}}\hat{q}\left[
\frac{\ln(2\hat{q})+1}{\hat{q}}-\frac{\ln(2\hat{b})+1}{\hat{b}}\right]
+\frac{A^{(1)}\beta_{2}}{8\beta_{1}^{3}}\left[
\ln^{2}(2\hat{q})-\ln^{2}(2\hat{b})\right],
\nonumber \\
&&+\frac{A^{(1)}\beta_{2}}{8\beta_{1}^{3}}
\ln\left(\frac{e^{2\gamma_E-1}}{2}\right)\left[
\frac{\ln(2\hat{q})+1}{\hat{q}}-\frac{\ln(2\hat{b})+1}{\hat{b}}\right]
-\frac{A^{(2)}\beta_{2}}{16\beta_{1}^{4}}\left[
\frac{2\ln(2\hat{q})+3}{\hat{q}}-\frac{2\ln(2\hat{b})+3}{\hat{b}}\right]
\nonumber \\
& &-\frac{A^{(2)}\beta_{2}}{16\beta_{1}^{4}}
\frac{\hat{q}-\hat{b}}{\hat{b}^2}\left[2\ln(2\hat{b})+1\right]
+\frac{A^{(2)}\beta_{2}^2}{432\beta_{1}^{6}}
\frac{\hat{q}-\hat{b}}{\hat{b}^3}
\left[9\ln^2(2\hat{b})+6\ln(2\hat{b})+2\right]
\nonumber \\
&& +\frac{A^{(2)}\beta_{2}^2}{1728\beta_{1}^{6}}\left[
\frac{18\ln^2(2\hat{q})+30\ln(2\hat{q})+19}{\hat{q}^2}
-\frac{18\ln^2(2\hat{b})+30\ln(2\hat{b})+19}{\hat{b}^2}\right],\label{sq}
\end{eqnarray}
with the variables
\begin{eqnarray}
\hat q\equiv \mbox{ln}[Q/(\sqrt 2\Lambda_{\rm QCD})],~~~ \hat b\equiv
\mbox{ln}[1/(b\Lambda_{\rm QCD})],
\end{eqnarray}
and the coefficients $A^{(i)}$ and $\beta_i$,
\begin{eqnarray}
\beta_1=\frac{33-2n_f}{12},~~\beta_2=\frac{153-19n_f}{24},\nonumber\\
A^{(1)}=\frac{4}{3},~~A^{(2)}=\frac{67}{9}
-\frac{\pi^2}{3}-\frac{10}{27}n_f+\frac{8}{3}\beta_1\mbox{ln}(\frac{1}{2}e^{\gamma_E}),
\end{eqnarray}
where $n_f$ is the number of the quark flavors and $\gamma_E$ is the Euler
constant. We adopt the one-loop running coupling constant, so
only the first four terms of Eq.~(\ref{sq}) are picked up in the numerical analysis.

\end{appendix}


\begin{thebibliography}{1}

\bibitem{Aoki:2019cca}
S.~Aoki \textit{et al.} [Flavour Lattice Averaging Group],
Eur. Phys. J. C \textbf{80}, no.2, 113 (2020)
doi:10.1140/epjc/s10052-019-7354-7
[arXiv:1902.08191 [hep-lat]].

\bibitem{Bediaga:2012py}
R.~Aaij \textit{et al.} [LHCb],
Eur. Phys. J. C \textbf{73}, no.4, 2373 (2013)
doi:10.1140/epjc/s10052-013-2373-2
[arXiv:1208.3355 [hep-ex]].

\bibitem{Kou:2018nap}
E.~Kou \textit{et al.} [Belle-II],
PTEP \textbf{2019}, no.12, 123C01 (2019)
[erratum: PTEP \textbf{2020}, no.2, 029201 (2020)]
doi:10.1093/ptep/ptz106
[arXiv:1808.10567 [hep-ex]].

\bibitem{Cerri:2018ypt}
A.~Cerri, V.~V.~Gligorov, S.~Malvezzi, J.~Martin Camalich, J.~Zupan, S.~Akar, J.~Alimena, B.~C.~Allanach, W.~Altmannshofer and L.~Anderlini, \textit{et al.}
CERN Yellow Rep. Monogr. \textbf{7}, 867-1158 (2019)
doi:10.23731/CYRM-2019-007.867
[arXiv:1812.07638 [hep-ph]].

\bibitem{Ball:2004ye}
P.~Ball and R.~Zwicky,
Phys. Rev. D \textbf{71}, 014015 (2005)
doi:10.1103/PhysRevD.71.014015
[arXiv:hep-ph/0406232 [hep-ph]].

\bibitem{Ball:2006wn}
  P.~Ball, V.~M.~Braun and A.~Lenz,
  JHEP {\bf 0605}, 004 (2006)
  doi:10.1088/1126-6708/2006/05/004
  [hep-ph/0603063].

\bibitem{Ball:2007rt}
P.~Ball and G.~W.~Jones,
JHEP \textbf{03}, 069 (2007)
doi:10.1088/1126-6708/2007/03/069
[arXiv:hep-ph/0702100 [hep-ph]].

\bibitem{Bali:2017ude}
  G.~S.~Bali {\it et al.} [RQCD Collaboration],
  Phys.\ Lett.\ B {\bf 774}, 91 (2017)
  doi:10.1016/j.physletb.2017.08.077
  [arXiv:1705.10236 [hep-lat]].

\bibitem{Bali:2019dqc}
  G.~S.~Bali {\it et al.} [RQCD Collaboration],
  JHEP {\bf 1908}, 065 (2019)
  Addendum: [JHEP {\bf 2011}, 037 (2020)]
  doi:10.1007/JHEP08(2019)065, 10.1007/JHEP11(2020)037
  [arXiv:1903.08038 [hep-lat]].

\bibitem{Hua:2020gnw}
  J.~Hua, M.~H.~Chu, P.~Sun, W.~Wang, J.~Xu, Y.~B.~Yang, J.~H.~Zhang and Q.~A.~Zhang,
  arXiv:2011.09788 [hep-lat].

\bibitem{Chernyak:1990ag}
V.~L.~Chernyak and I.~R.~Zhitnitsky,
Nucl. Phys. B \textbf{345}, 137-172 (1990)
doi:10.1016/0550-3213(90)90612-H

\bibitem{Ali:1993vd}
A.~Ali, V.~M.~Braun and H.~Simma,
Z. Phys. C \textbf{63}, 437-454 (1994)
doi:10.1007/BF01580324
[arXiv:hep-ph/9401277 [hep-ph]].

\bibitem{Khodjamirian:2000mi}
A.~Khodjamirian,
Nucl. Phys. B \textbf{605}, 558-578 (2001)
doi:10.1016/S0550-3213(01)00194-8
[arXiv:hep-ph/0012271 [hep-ph]].

\bibitem{Beneke:1999br}
M.~Beneke, G.~Buchalla, M.~Neubert and C.~T.~Sachrajda,
Phys. Rev. Lett. \textbf{83}, 1914-1917 (1999)
doi:10.1103/PhysRevLett.83.1914
[arXiv:hep-ph/9905312 [hep-ph]].

\bibitem{Li:1994cka}
H.~n.~Li and H.~L.~Yu,
Phys. Rev. Lett. \textbf{74}, 4388-4391 (1995)
doi:10.1103/PhysRevLett.74.4388
[arXiv:hep-ph/9409313 [hep-ph]].

\bibitem{Li:1995jr}
H.~n.~Li and H.~L.~Yu,
Phys. Lett. B \textbf{353}, 301-305 (1995)
doi:10.1016/0370-2693(95)00557-2

\bibitem{Li:1994iu}
H.~n.~Li and H.~L.~Yu,
Phys. Rev. D \textbf{53}, 2480-2490 (1996)
doi:10.1103/PhysRevD.53.2480
[arXiv:hep-ph/9411308 [hep-ph]].

\bibitem{Keum:2000ms}
Y.~Y.~Keum and H.~n.~Li,
Phys. Rev. D \textbf{63}, 074006 (2001)
doi:10.1103/PhysRevD.63.074006
[arXiv:hep-ph/0006001 [hep-ph]].
\bibitem{Keum:2000wi}
Y.~Y.~Keum, H.~n.~Li and A.~I.~Sanda,
Phys. Rev. D \textbf{63}, 054008 (2001)
doi:10.1103/PhysRevD.63.054008
[arXiv:hep-ph/0004173 [hep-ph]].

\bibitem{Lu:2000em}
C.~D.~Lu, K.~Ukai and M.~Z.~Yang,
Phys. Rev. D \textbf{63}, 074009 (2001)
doi:10.1103/PhysRevD.63.074009
[arXiv:hep-ph/0004213 [hep-ph]].

\bibitem{Bauer:2000ew}
C.~W.~Bauer, S.~Fleming and M.~E.~Luke,
Phys. Rev. D \textbf{63}, 014006 (2000)
doi:10.1103/PhysRevD.63.014006
[arXiv:hep-ph/0005275 [hep-ph]].

\bibitem{Bauer:2000yr}
C.~W.~Bauer, S.~Fleming, D.~Pirjol and I.~W.~Stewart,
Phys. Rev. D \textbf{63}, 114020 (2001)
doi:10.1103/PhysRevD.63.114020
[arXiv:hep-ph/0011336 [hep-ph]].
\bibitem{Li:2020rcg}
H.~D.~Li, C.~D.~L\"u, C.~Wang, Y.~M.~Wang and Y.~B.~Wei,
JHEP \textbf{04}, 023 (2020)
doi:10.1007/JHEP04(2020)023
[arXiv:2002.03825 [hep-ph]].

\bibitem{Huber:2016xod}
T.~Huber, S.~Kr\"ankl and X.~Q.~Li,
JHEP \textbf{09}, 112 (2016)
doi:10.1007/JHEP09(2016)112
[arXiv:1606.02888 [hep-ph]].

\bibitem{Bell:2020qus}
G.~Bell, M.~Beneke, T.~Huber and X.~Q.~Li,
JHEP \textbf{04}, 055 (2020)
doi:10.1007/JHEP04(2020)055
[arXiv:2002.03262 [hep-ph]].


\bibitem{Zyla:2020zbs}
P.~A.~Zyla \textit{et al.} [Particle Data Group],
PTEP \textbf{2020}, no.8, 083C01 (2020)
doi:10.1093/ptep/ptaa104

\bibitem{Jones:1999rz}
M.~K.~Jones \textit{et al.} [Jefferson Lab Hall A],
Phys. Rev. Lett. \textbf{84}, 1398-1402 (2000)
doi:10.1103/PhysRevLett.84.1398
[arXiv:nucl-ex/9910005 [nucl-ex]].

\bibitem{Gayou:2001qd}
O.~Gayou \textit{et al.} [Jefferson Lab Hall A],
Phys. Rev. Lett. \textbf{88}, 092301 (2002)
doi:10.1103/PhysRevLett.88.092301
[arXiv:nucl-ex/0111010 [nucl-ex]].

\bibitem{Ralston:2003mt}
J.~P.~Ralston and P.~Jain,
Phys. Rev. D \textbf{69}, 053008 (2004)
doi:10.1103/PhysRevD.69.053008
[arXiv:hep-ph/0302043 [hep-ph]].

\bibitem{Nandi:2007qx}
S.~Nandi and H.~n.~Li,
Phys. Rev. D \textbf{76}, 034008 (2007)
doi:10.1103/PhysRevD.76.034008
[arXiv:0704.3790 [hep-ph]].

\bibitem{Botts:1989kf}
J.~Botts and G.~F.~Sterman,
Nucl. Phys. B \textbf{325}, 62-100 (1989)
doi:10.1016/0550-3213(89)90372-6

\bibitem{Li:1992nu}
H.~n.~Li and G.~F.~Sterman,
Nucl. Phys. B \textbf{381}, 129-140 (1992)
doi:10.1016/0550-3213(92)90643-P


\bibitem{Li:2001ye}
H.~n.~Li,
Nucl. Phys. A \textbf{684}, 304-306 (2001)
doi:10.1016/S0375-9474(01)00493-6

\bibitem{Szczepaniak:1990dt}
A.~Szczepaniak, E.~M.~Henley and S.~J.~Brodsky,
Phys. Lett. B \textbf{243}, 287-292 (1990)
doi:10.1016/0370-2693(90)90853-X

\bibitem{Burdman:1992hg}
G.~Burdman and J.~F.~Donoghue,
Phys. Lett. B \textbf{270}, 55-60 (1991)
doi:10.1016/0370-2693(91)91538-7

\bibitem{Beneke:2000wa}
M.~Beneke and T.~Feldmann,
Nucl. Phys. B \textbf{592}, 3-34 (2001)
doi:10.1016/S0550-3213(00)00585-X
[arXiv:hep-ph/0008255 [hep-ph]].

\bibitem{Collins:1981uk}
J.~C.~Collins and D.~E.~Soper,
Nucl. Phys. B \textbf{193}, 381 (1981)
[erratum: Nucl. Phys. B \textbf{213}, 545 (1983)]
doi:10.1016/0550-3213(81)90339-4

\bibitem{Kurimoto:2001zj}
T.~Kurimoto, H.~n.~Li and A.~I.~Sanda,
Phys. Rev. D \textbf{65}, 014007 (2002)
doi:10.1103/PhysRevD.65.014007
[arXiv:hep-ph/0105003 [hep-ph]].

\bibitem{Wei:2002iu}
Z.~T.~Wei and M.~Z.~Yang,
Nucl. Phys. B \textbf{642}, 263-289 (2002)
doi:10.1016/S0550-3213(02)00623-5
[arXiv:hep-ph/0202018 [hep-ph]].

\bibitem{Akhoury:1993uw}
R.~Akhoury, G.~F.~Sterman and Y.~P.~Yao,
Phys. Rev. D \textbf{50}, 358-372 (1994)
doi:10.1103/PhysRevD.50.358

\bibitem{Korchemsky:1999qb}
G.~P.~Korchemsky, D.~Pirjol and T.~M.~Yan,
Phys. Rev. D \textbf{61}, 114510 (2000)
doi:10.1103/PhysRevD.61.114510
[arXiv:hep-ph/9911427 [hep-ph]].

\bibitem{Li:2001ay}
H.~n.~Li,
Phys. Rev. D \textbf{66}, 094010 (2002)
doi:10.1103/PhysRevD.66.094010
[arXiv:hep-ph/0102013 [hep-ph]].


\bibitem{Zhang:2020qaz}
Z.~Q.~Zhang and H.~n.~Li,
[arXiv:2007.11173 [hep-ph]].


\bibitem{Li:2013xna}
H.~n.~Li, Y.~L.~Shen and Y.~M.~Wang,
JHEP \textbf{01}, 004 (2014)
doi:10.1007/JHEP01(2014)004
[arXiv:1310.3672 [hep-ph]].

\bibitem{Beneke:2000ry}
M.~Beneke, G.~Buchalla, M.~Neubert and C.~T.~Sachrajda,
Nucl. Phys. B \textbf{591}, 313-418 (2000)
doi:10.1016/S0550-3213(00)00559-9
[arXiv:hep-ph/0006124 [hep-ph]].

\bibitem{Beneke:2001ev}
M.~Beneke, G.~Buchalla, M.~Neubert and C.~T.~Sachrajda,
Nucl. Phys. B \textbf{606}, 245-321 (2001)
doi:10.1016/S0550-3213(01)00251-6
[arXiv:hep-ph/0104110 [hep-ph]].

\bibitem{Chang:1996dw}
C.~H.~V.~Chang and H.~n.~Li,
Phys. Rev. D \textbf{55}, 5577-5580 (1997)
doi:10.1103/PhysRevD.55.5577
[arXiv:hep-ph/9607214 [hep-ph]].

\bibitem{Yeh:1997rq}
T.~W.~Yeh and H.~n.~Li,
Phys. Rev. D \textbf{56}, 1615-1631 (1997)
doi:10.1103/PhysRevD.56.1615
[arXiv:hep-ph/9701233 [hep-ph]].

\bibitem{Cheng:1999gs}
H.~Y.~Cheng, H.~n.~Li and K.~C.~Yang,
Phys. Rev. D \textbf{60}, 094005 (1999)
doi:10.1103/PhysRevD.60.094005
[arXiv:hep-ph/9902239 [hep-ph]].

\bibitem{Li:2001vm}
H.~n.~Li,
[arXiv:hep-ph/0110365 [hep-ph]].

\bibitem{Ali:2007ff}
A.~Ali, G.~Kramer, Y.~Li, C.~D.~Lu, Y.~L.~Shen, W.~Wang and Y.~M.~Wang,
Phys. Rev. D \textbf{76}, 074018 (2007)
doi:10.1103/PhysRevD.76.074018
[arXiv:hep-ph/0703162 [hep-ph]].


\bibitem{Peter:2020}
P.~Lepage and C.~Gohlke, gplepage/lsqfit: lsqfit version 11.7,  Zenodo. http://doi.org/10.5281/zenodo.4037174

\bibitem{Aaij:2020wnj}
R.~Aaij {\it et al.} [LHCb Collaboration],
arXiv:2012.12789 [hep-ex].

\bibitem{Li:2006jv}
H.~n.~Li and S.~Mishima,
Phys. Rev. D \textbf{74}, 094020 (2006)
doi:10.1103/PhysRevD.74.094020
[arXiv:hep-ph/0608277 [hep-ph]].

\bibitem{Rui:2011dr}
Z.~Rui, X.~Gao and C.~D.~Lu,
Eur. Phys. J. C \textbf{72}, 1923 (2012)
doi:10.1140/epjc/s10052-012-1923-3
[arXiv:1111.0181 [hep-ph]].



\bibitem{Li:2012bp}
H.~n.~Li, Y.~L.~Shen and Y.~M.~Wang,
Phys. Rev. D \textbf{85}, 074004 (2012)
doi:10.1103/PhysRevD.85.074004

\bibitem{Wang:2012ab}
W.~F.~Wang and Z.~J.~Xiao,
Phys. Rev. D \textbf{86}, 114025 (2012)
doi:10.1103/PhysRevD.86.114025
[arXiv:1207.0265 [hep-ph]].

\bibitem{Yang:2020xal}
Y.~Yang, L.~Lang, X.~Zhao, J.~Huang and J.~Sun,
[arXiv:2012.10581 [hep-ph]].

\bibitem{Lu:2000hj}
C.~D.~Lu and M.~Z.~Yang,
Eur. Phys. J. C \textbf{23}, 275-287 (2002)
doi:10.1007/s100520100878
[arXiv:hep-ph/0011238 [hep-ph]].

\bibitem{Zhou:2019crd}
S.~H.~Zhou and C.~D.~L\"u,
Chin. Phys. C \textbf{44}, no.6, 063101 (2020)
doi:10.1088/1674-1137/44/6/063101
[arXiv:1910.03160 [hep-ph]].

\bibitem{Cheng:2020hyj}
H.~Y.~Cheng,
[arXiv:2005.06080 [hep-ph]].

\bibitem{Li:2016tpn}
Y.~Li, A.~J.~Ma, W.~F.~Wang and Z.~J.~Xiao,
Phys. Rev. D \textbf{95}, no.5, 056008 (2017)
doi:10.1103/PhysRevD.95.056008
[arXiv:1612.05934 [hep-ph]].

\bibitem{Yan:2017nlj}
  D.~C.~Yan, P.~Yang, X.~Liu and Z.~J.~Xiao,
  Nucl.\ Phys.\ B {\bf 931}, 79 (2018)
  doi:10.1016/j.nuclphysb.2018.04.007
  [arXiv:1707.06043 [hep-ph]].

\bibitem{Wang:2017ijn}
  Y.~M.~Wang and Y.~L.~Shen,
  JHEP {\bf 1712}, 037 (2017)
  doi:10.1007/JHEP12(2017)037
  [arXiv:1706.05680 [hep-ph]].

\bibitem{Wang:2018wfj}
  Y.~M.~Wang and Y.~L.~Shen,
  JHEP {\bf 1805}, 184 (2018)
  doi:10.1007/JHEP05(2018)184
  [arXiv:1803.06667 [hep-ph]].


\end{thebibliography}
\end{document}